\documentclass[preprint,showpacs,preprintnumbers,
amsmath,amssymb]{revtex4}

\usepackage{graphicx}
\usepackage{dcolumn}
\usepackage{bm,morefloats,color}

\usepackage{amssymb,bm}

\def\Eq#1{\begin{equation} #1 \end{equation}}
\def\Eqr#1{\begin{eqnarray} #1 \end{eqnarray}}
\def\Eqrsubl#1#2{\begin{subequations}\label{#1}\Eqr{#2}\end{subequations}}

\newcommand{\nn}{\nonumber}
\newcommand{\pd}{\partial}

\newcommand{\bea}{\begin{eqnarray}}
\newcommand{\eea}{\end{eqnarray}}

\def\Xsp{{\rm X}}

\def\Ysp{{\rm Y}}

\def\X5sp{{\rm X}_5}
\def\Y3sp{{\rm Y}_3}
\def\Z3sp{{\rm Z}_3}

\def\lap{{\triangle}}
\def\e{{\rm e}}

\begin{document}

\title{Violation of cosmic censorship in dynamical $p$-brane systems}

\author{Kengo Maeda}%
\affiliation{%
Faculty of Engineering, 
Shibaura Institute of Technology, 
Saitama 330-8570, Japan}%

\author{Kunihito Uzawa}
\affiliation{%
Department of Physics,
School of Science and Technology,
Kwansei Gakuin University, Sanda, Hyogo 669-1337, Japan
}%

\date{\today}

\begin{abstract}
We study the cosmic censorship of dynamical $p$-brane systems in a 
$D$-dimensional background. 
This is the generalization of the analysis 
in the Einstein-Maxwell-dilaton theory, which was discussed by Horne and 
Horowitz [Phys. Rev. D 48, R5457 (1993)].
We show that a timelike curvature singularity generically appears 
from an asymptotic region in the time evolution of the $p$-brane solution. 
Since we can set regular and smooth initial data in 
a dynamical M5-brane system in 11-dimensional supergravity, 
this implies a violation of cosmic censorship. 
\end{abstract}


\maketitle


\section{Introduction}
\label{sec:introduction}

The dynamical $p$-brane solutions in gravity theory were introduced 
in \cite{Gibbons:2005rt, Chen:2005jp, Kodama:2005fz, Kodama:2005cz, 
Kodama:2006ay, Binetruy:2007tu, Binetruy:2008ev, Maeda:2009tq, 
Maeda:2009zi, Gibbons:2009dr, Maeda:2009ds, Uzawa:2010zza, Maeda:2010ja, 
Minamitsuji:2010fp, Maeda:2010aj, Minamitsuji:2010kb, Uzawa:2010zz, 
Nozawa:2010zg, Minamitsuji:2010uz, Maeda:2011sh, Minamitsuji:2011jt, 
Maeda:2012xb, Blaback:2012mu, Minamitsuji:2012if, Uzawa:2013koa, 
Uzawa:2013msa, Blaback:2013taa, Uzawa:2014kka, Uzawa:2014dra} 
and have been widely used ever since. 
However, some aspects of the physical properties, 
such as curvature singularities and 
causal structure, have remained slightly unclear. 
In the present paper, 
we investigate the cosmic censorship conjecture, aiming to give simpler 
and more direct demonstrations of the $p$-brane dynamics near curvature 
singularities. Our work is the generalization of analysis for 
the cosmic censorship in Einstein-Maxwell-dilaton theory, 
which was given by Horne and Horowitz \cite{Horne:1993sy}.

The singularities on spacetime that 
appear due to the existence of time dependence 
in Einstein-Maxwell theories were studied from several points of 
view in \cite{Kastor:1992nn, Maki:1992tq}. 
In Refs.~\cite{Brill:1993tm, Horne:1993sy}, 
a violation of cosmic censorship \cite{Penrose:1964wq, 
Penrose:1969pc} has been discussed. 
The other completely consistent calculation for spacetime singularity in 
Einstein-Maxwell theories was performed in 
\cite{Maki:1994wc}. 

In Sec.~\ref{sec:db}, we will review the dynamical $p$-brane solutions, 
which have dilaton coupling constants with the same value as 
static $p$-brane backgrounds \cite{Gibbons:2005rt, 
Binetruy:2007tu, Binetruy:2008ev, Maeda:2009zi, 
Minamitsuji:2010kb, Minamitsuji:2010uz, Uzawa:2014dra}. 
This is a good illustration to describe $p$-branes in an expanding universe 
and also an important example  
to see the naked singularity of a particular type. 
These singularities also appear in a variety of D$p$-brane configurations 
of string theory. 

We analyze the null geodesics 
 in dynamical $p$-brane backgrounds 
and study the causal structure of a dynamical 
$p$-brane in detail to examine the cosmic censorship conjecture 
in Sec.~\ref{sec:gp}.  
We verify that the dynamical $p$-brane backgrounds generically possess 
a timelike curvature singularity 
whose location is far away from the origin of the $p$-brane. 
This implies a violation of cosmic censorship. 
For a vanishing or trivial dilaton such as the dynamical M-brane and D3-brane 
system in supergravity, 
the regular initial data evolve into a spacetime with a 
naked singularity. Then, we have a violation of cosmic censorship. 
We will see it for the dynamical M5-brane solution in the 11-dimensional 
supergravity. 

We will discuss that the presence of a $p$-brane 
drastically 
changes the causal structure 
of the dynamical solution. In particular, it turns out that in 
the absence of any $p$-branes, the solution describes a collapsing 
Friedmann-Lema\^itre-Robertson-Walker (FLRW) universe 
with a final spacelike or null singularities.
If only one single $p$-brane is added, 
these singularities are replaced by a timelike or a null one. 
Section \ref{sec:discussions} 
is devoted to summary and discussion. 

In Appendix \ref{sec:ap}, we will review the explicit form of 
the components of Einstein equations. 
We will also give the global structure of a static $p$-brane in Appendix 
\ref{sec:static}. 
In Appendix \ref{sec:b}, we extend the 
analysis to include the "massive brane". 
We provide more details about the Killing symmetries 
of the solution in the presence of a massive brane background 
that are obtained in type IIA string theory.  


\section{Dynamical brane backgrounds}
\label{sec:db}

In this section, 
we will review the dynamical brane systems
in $D$ dimensions. 
We consider a $D$-dimensional theory which is 
composed of the metric $g_{MN}$, the scalar field $\phi$, 
and antisymmetric tensor field strength of rank $(p+2)$. 
The action in $D$ dimensions is 
given by 
\Eqr{
S=\frac{1}{2\kappa^2}\int \left[R\,\ast{\bf 1}_D
 -\frac{1}{2}\ast d\phi \wedge d\phi
 -\frac{1}{2}\frac{1}{\left(p+2\right)!}
 \e^{\epsilon c\phi}\ast F_{(p+2)}\,\wedge\, 
 F_{\left(p+2\right)} \right],
\label{p:a:Eq}
}
where $R$ denotes the Ricci scalar 
with respect to the $D$-dimensional metric $g_{MN}$\,, 
$\kappa^2$ is the $D$-dimensional gravitational constant, 
$\ast$ denotes the Hodge operator in the $D$-dimensional spacetime, and 
$F_{\left(p+2\right)}$ is $\left(p+2\right)$-form field strength, respectively.
The constants $c$\,, $\epsilon$ are defined by 
\Eqrsubl{p:parameters:Eq}{
c^2&=&4-\frac{2(p+1)(D-p-3)}{D-2},
   \label{p:c:Eq}\\
\epsilon&=&\left\{
\begin{array}{cc}
 +&{\rm if~the}~p-{\rm brane~is~electric}\\
 -&~~~{\rm if~the}~p-{\rm brane~is~magnetic}\,,
\end{array} \right.
 \label{p:epsilon:Eq}
   }
respectively. 
This is part of the low energy action from string 
theory in Einstein frame. 
The $(p+2)$-form field strength $F_{(p+2)}$ is written by the 
$\left(p+1\right)$-form gauge potential $A_{\left(p+1\right)}$, 
respectively:
\Eq{
F_{(p+2)}=dA_{(p+1)}\,.
}
The solution of dynamical $p$-brane is given by 
\cite{Binetruy:2007tu, Binetruy:2008ev, Maeda:2009zi, 
Uzawa:2010zza, Minamitsuji:2010kb, Minamitsuji:2010uz, Uzawa:2014dra}
\Eqrsubl{p:ansatz:Eq}{
ds^2&=&h^a(x, y)\,\eta_{\mu\nu}(\Xsp)dx^{\mu}dx^{\nu}+h^b(x, y)\,\delta_{ij}
(\Ysp)dy^idy^j\,,
  \label{p:metric:Eq}\\
\e^{\phi}&=&h^{\epsilon c/2}\,,
  \label{p:dilaton:Eq}\\
F_{\left(p+2\right)}&=&
d\left[h^{-1}(x, y)\right]\wedge
\,dx^0\,\wedge\,dx^1\wedge\,\cdots\,\wedge\,dx^p\,,
  \label{p:strength:Eq}
}
where $\eta_{\mu\nu}$ is the $(p+1)$-dimensional Minkowski metric, 
 $\delta_{ij}$ is $(D-p-1)$-dimensional Euclidean metric,
respectively, the parameters $a$, 
$b$ in the metric (\ref{p:metric:Eq}) are given by 
\Eq{
a=-\frac{D-p-3}{D-2}\,,~~~~~~~~~
b=\frac{p+1}{D-2}\,.
 \label{p:parameter:Eq}
}
The function $h(x, y)$ in the $D$-dimensional metric (\ref{p:metric:Eq})  
can be obtained explicitly as 
\Eqrsubl{pv:warp:Eq}{ 
h(x, y)&=&h_0(x)+h_1(y)\,,\\
h_0(x)&=&c_\mu x^\mu + \bar{c}\,,
  \label{pv:h0:Eq}\\
h_1(y)&=&1+\sum_l\frac{M_l}{|y^i-y^i_l|^{D-p-3}}\,, ~~~~
{\rm for}~~~D-p-3\ne 0\,,
  \label{pv:h1:Eq}\\
h_1(y)&=&1+\sum_lM_l\ln |y^i-y^i_l|\,, ~~~~
{\rm for}~~~D-p-3=0\,,
  \label{pv:h2:Eq}
}
where $c_\mu$, $\bar{c}$, $M_l$ and $y^i_l$ 
are constant parameters.

There is a possibility of smearing out
some dimensions of Y space. 
The function 
$h_1$ in (\ref{pv:h1:Eq}) can be expressed as \cite{Maeda:2012xb}
\Eqrsubl{pv:smear:Eq}{
h_1(y)&=&1+\sum_l\frac{M_l}{|y^i-y^i_l|^{D-p-3-d_\Ysp}}\,, ~~~~
{\rm for}~~~d_\Ysp\ne D-p-3\,,
  \label{pv:h1-1:Eq}\\
h_1(y)&=&1+\sum_lM_l\ln |y^i-y^i_l|\,, ~~~~
{\rm for}~~~d_\Ysp = D-p-3\,,
  \label{pv:h1-2:Eq}
}
where $d_\Ysp$ is the smeared dimensions in Y space.

\section{Geometry of the dynamical $p$-brane}
\label{sec:gp}

Now we will study the spacetime structure of the dynamical $p$-brane 
backgrounds. 
The scalar curvature of the metric (\ref{p:metric:Eq}) becomes 
\Eqr{
R=-\frac{c^2}{8}
h^{-a}\eta^{\rho\sigma}\pd_{\rho}\ln h\pd_{\sigma}\ln h
+\left[\frac{1}{8}\left(c^2-4\right)+b\right]h^{-b}
\delta^{kl}\pd_k\ln h\pd_l\ln h\,,
\label{n:bM:Eq}
}
where the function $h$ is given by (\ref{pv:warp:Eq}). 
Since the Ricci scalar diverges at $h=0$, there are curvature 
singularities at these spacetime points.  The singularity is timelike and 
its effects can propagate into the background spacetime. 
The naked singularity thus appears 
at $|y^i-y^i_l|\rightarrow\infty$\,, 
and moves in toward $y^i\rightarrow y^i_l$\,,  
in the $p$-brane background (\ref{p:ansatz:Eq}).  
Setting $h_0=c_0\,t$\,, and $c_0<0$\,, 
the region of regular spacetime eventually splits, 
and surrounds each of the $p$-branes separately as $t$ increases. 
These are the same results as in the background \cite{Brill:1993tm}. 
For $p$-brane with $c\ne 0$, the Ricci scalar curvature 
also diverges at $y^i=y^i_l$\,. 

In this section, we discuss the behavior of geometrical structure 
near the curvature singularities. The dynamical $p$-brane backgrounds indeed 
have singularities at $y^i=y^i_l$ and $h=0$ if there is non-trivial 
dilaton. Although the former  
appears in the static $p$-brane solution as well as dynamical one, 
the latter is the singularities 
caused by the gravitational 
collapse in the dynamical $p$-brane background. 
First we comment about causal structure and 
a curvature singularity at $y^i=y^i_l$. 
The dynamical $p$-brane solutions in this paper describe 
extremal black holes where a non-spacelike singularity 
appears at each location of the dynamical $p$-branes, $y^i=y^i_l$. 
Although the singularity at $y^i=y^i_l$ is not hidden by an event horizon,
this will be removed in the non-extremal solution \cite{Horne:1993sy}, or 
string theory \cite{Johnson:1999qt, Jarv:2000zv, Peet:1999nh, Johnson:2000ch, 
Yamaguchi:2001yd}. 
Hence, the curvature singularity at $y^i=y^i_l$ is not so physically 
significant because of the expectation of the singularity resolutions.  
In this paper, we will set the "regular and smooth" initial data and 
focus on the behavior of curvature singularities at $h=0$\,,  
even if the initial data at $y^i=y^i_l$ describes a singularity. 
Here, the "regular and smooth" means that we can set an initial 
data describing 
 smooth universe except for the location of the $p$-brane. 
The "regular and smooth" initial data in the dynamical 
$p$-brane background (\ref{p:metric:Eq}) evolves into a 
spacetime with a naked singularity. 

If the dilaton is trivial or vanishing, we can find the horizon at 
$y^i=y^i_l$ for the dynamical brane solution \cite{Maeda:2009zi}. 
For example, 
we obtain a black hole geometry for D3-brane, M-brane system 
in the ten-, or eleven-dimensional supergravity.  
Although the near horizon geometries of these black holes in the expanding
universe are the same as the static solutions, 
the asymptotic structures are completely different, giving the FLRW 
universe with scale factors same as the universe filled by stiff matter
\cite{Maeda:2009zi}. 
In this case, we will be able to set a perfectly smooth initial data. 
When it evolves into a timelike curvature singularity, 
the cosmic censorship will be violated. 
We will see this issues in sec.~\ref{sec:singular}. 

Let us study the null geodesic in order to discuss the property of 
the singularity in the dynamical $p$-brane background (\ref{pv:warp:Eq}). 
We set 
the constant vector $c_A x^A=c_1 x^1~(A=1\,,\cdots\,, p)$ 
after rotating the spatial part of worldvolume coordinates on 
$t=$constant surface. 
If we boost in the $x^1$-direction, 
then the function $h_0$ depends on the coordinate of 
(i) time~($|c_1|<|c_0|$), (ii) null~($|c_1|=|c_0|$), 
(iii) spatial part of worldvolume~($|c_1|>|c_0|$). 
In the following, we will discuss two cases (i) in sec.~\ref{sec:singular} 
and (ii) in sec.~\ref{sec:null}. 

It is useful to comment about the Killing symmetries of the solution. 
It depends on $x^\mu$ and $y^i$ through the combination 
$h(x, y)=h_0(x)+h_1(y)$\,. 
Since the Killing vectors $\xi^{\mu}~(\mu=0\,,\cdots\,, p)$ 
of the dynamical $p$-brane background satisfy $\xi^\mu c_\mu=0$\,,  
there are only $p$ independent vectors for $\xi^{\mu}$\,.

%
%
\subsection{The singularities on time-dependent 
$p$-brane background} 
\label{sec:singular}
Now we discuss the causal structure of a dynamical $p$-brane background 
where the function $h_0$ depends only on the time. The solution can be 
expressed as 
\Eqrsubl{ge:bg:Eq}{
&&ds^2=h^a(t, r)\eta_{\mu\nu}(\Xsp)dx^{\mu}dx^{\nu}+h^b(t, r)\delta_{ij}
(\Ysp)dy^idy^j\,,
 \label{g:metric:Eq}\\
&&\delta_{ij}(\Ysp)dy^idy^j=dr^2+r^2w_{mn}dz^mdz^n\,,\\
&&h(t, r)=h_0(t)+h_1(r),~~~~~h_0(t)=c_0t+\bar{c}
\,,~~~~~
h_1(r)=\frac{M}{r^{D-p-3}}\,, 
  \label{g:h:Eq}
}
where we assume $D-p-3>0$\,, $d_\Ysp=0$\,, $\eta_{\mu\nu}(\Xsp)$\,, 
$w_{mn}$\,, 
are the metrics of $(p+1)$-dimensional Minkowski spacetime, 
$(D-p-2)$-dimensional sphere, respectively, 
$M$, $c_0$, $\bar{c}$
 are constant parameters, 
and the constants $a$, $b$ are given by (\ref{p:parameter:Eq}). 
Note that $\bar{c}$ can always be removed by 
shifting time as $t\rightarrow t-\bar{c}\,c_0^{-1}$ , 
up to the redefinition of $t$. So, we will set $\bar{c}=0$ henceforth. 
We discuss that future null geodesics arrive at the singularity in the 
finite affine parameter $s$. Now we consider the null geodesics 
with $\dot{z}^m=0$\,, 
where a dot denotes the derivative with respect to an affine parameter.
Then the geodesics in the $D$-dimensional spacetime 
(\ref{g:metric:Eq}) have to obey 
\Eq{
\dot{t}=\pm h^{1/2}\dot{r}\,,
   \label{g:tr:Eq}
}
where the $\pm$ denotes for future, past directed outgoing null 
geodesics, respectively. 
The geodesic equations for $t$ and $r$ give
\Eqrsubl{g:ge:Eq}{
&&\ddot{t}+\frac{1}{2}\left[\left(a+b\right)\pd_t\ln h
\pm 2a\,h^{-1/2}\pd_r \ln h\right]\,\dot{t}^2=0\,,
   \label{ge:ge-t:Eq}\\
&&\ddot{r}+\frac{1}{2}\left[\left(a+b\right)\pd_r\ln h
\pm 2b\,h^{1/2}\pd_t \ln h\right]\,\dot{r}^2=0\,.
   \label{ge:ge-r:Eq}
}

Since it is not easy to solve analytically, 
we first discuss the radial null geodesics without $p$-brane.
It turns out that the collapsing universe (\ref{ge:bg:Eq}) 
does not create any naked singularities. 

When $M=0$, the metric (\ref{ge:bg:Eq}) becomes
\Eq{
ds^2=\left(c_0\,t\right)^a\eta_{\mu\nu}(\Xsp)dx^{\mu}dx^{\nu}
+\left(c_0\,t\right)^b\left(dr^2+r^2w_{mn}dz^mdz^n\right)\,.
  \label{ge:M0:Eq}
}
If we define a new coordinate
\Eq{
\left(\frac{\tau}{\tau_0}\right)=(c_0\,t)^{\frac{a+2}{2}}\,,~~~~~
\tau_0=\frac{2}{c_0(a+2)}\,,
}
 the metric now takes the form
\Eq{
ds^2=
-d\tau^2+\left(\frac{\tau}{\tau_0}\right)^{\frac{2a}{a+2}}\delta_{AB}dx^Adx^B
+\left(\frac{\tau}{\tau_0}\right)^{\frac{2b}{a+2}}
\left(dr^2+r^2w_{mn}dz^mdz^n\right)\,,
}
where the metric $\delta_{AB}$ is the spatial part of the $p$-dimensional
Minkowski metric $\eta_{\mu\nu}$\,. 
Since this is simply a collapsing or an expanding 
FLRW universe, 
there is curvature singularity at $\tau=t=0$\,.

We next consider the time evolution of universe near the singularity.  
The behavior of singularity can be analyzed by
recalling that the solution of null geodesic equation 
 for $M=0$ case. 

From (\ref{ge:ge-t:Eq}), we find 
\Eq{
\ddot{t}+\frac{a+b}{2t}\,\dot{t}^2=0\,,
    \label{ge:ge-t2:Eq}
}
where the parameter $a+b$ satisfies the inequality $a+b> -1$ 
because of $p\ge 0$\,, $D\ge 4$\,. 
Then we find
\Eq{
t(s)=\beta_2\left[(1+\beta)s 
+\beta_1\right]^{\frac{1}{1+\beta}}\,, 
   \label{ge:t0:Eq}
}
where $\beta_i~(i=1\,, 2)$ denote constants, and $\beta$ is defined by
\Eq{
\beta=\frac{1}{2}\left(a+b\right)\,>-\frac{1}{2}\,.
 \label{ge:beta:Eq}
}
By Eq.~(\ref{ge:t0:Eq}), the null geodesics hit 
the curvature singularity 
at $t=0$ in finite affine parameter.  
Hence, the background geometry is 
geodesically 
incomplete. 

The solution of radial null geodesic equation can be also expressed as 
\Eq{
r(s)=\pm \,b^{-1}\,\sqrt{\frac{\beta_2}{c_0}}\left(2\beta+1\right)
\,\left[\left(1+\beta\right)s +\beta_1\right]^{\frac{1}{2(1+\beta)}}+r_0\,, 
  \label{ge:r1:Eq}
}
where $\beta_i~(i=1\,, 2)$ is given in (\ref{ge:t0:Eq}), $r_0$ is 
constant, and $\beta$ 
is defined by (\ref{ge:beta:Eq}). 

\begin{figure}[h]
 \begin{center}
\includegraphics[keepaspectratio, scale=0.29, angle=-90]{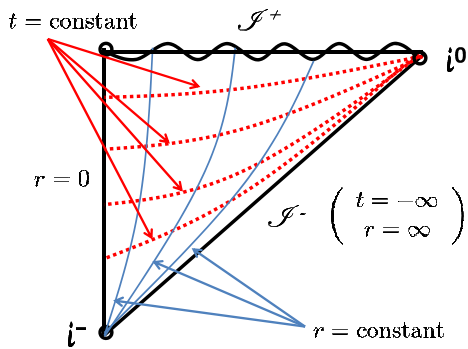}
  \caption{\baselineskip 14pt 
The Penrose diagram of the $D$-dimensional spacetime 
(\ref{g:h:Eq}) with $c_0=-1$\,, $t\le 0$\,,  
$M=0$\,, is depicted.  
The wavy line at $t=0$ is the spacelike singularity. 
}
  \label{fig:p}
 \end{center}
\end{figure}
The Penrose diagram for the spacetime with $M=0$ is shown in Fig.~\ref{fig:p}. 
The angular coordinates on $S^{D-p-2}$ and also $p$-dimensional 
worldvolume coordinates have been suppressed. 
The spacetime is geodesically incomplete, 
since radial null geodesics 
reach singularities at a finite affine parameter value. 

Although there is a curvature singularity at $t=0$\,, 
the singularity in this solution is spacelike. 
One way to see this is to note that outgoing radial light rays satisfy 
(\ref{ge:r1:Eq}), so that as $t\rightarrow 0$, $r$ approaches $r_0$\,, 
as $1+\beta>0$\,. 

\begin{figure}[h]
 \begin{center}
\includegraphics[keepaspectratio, scale=0.21, angle=0]{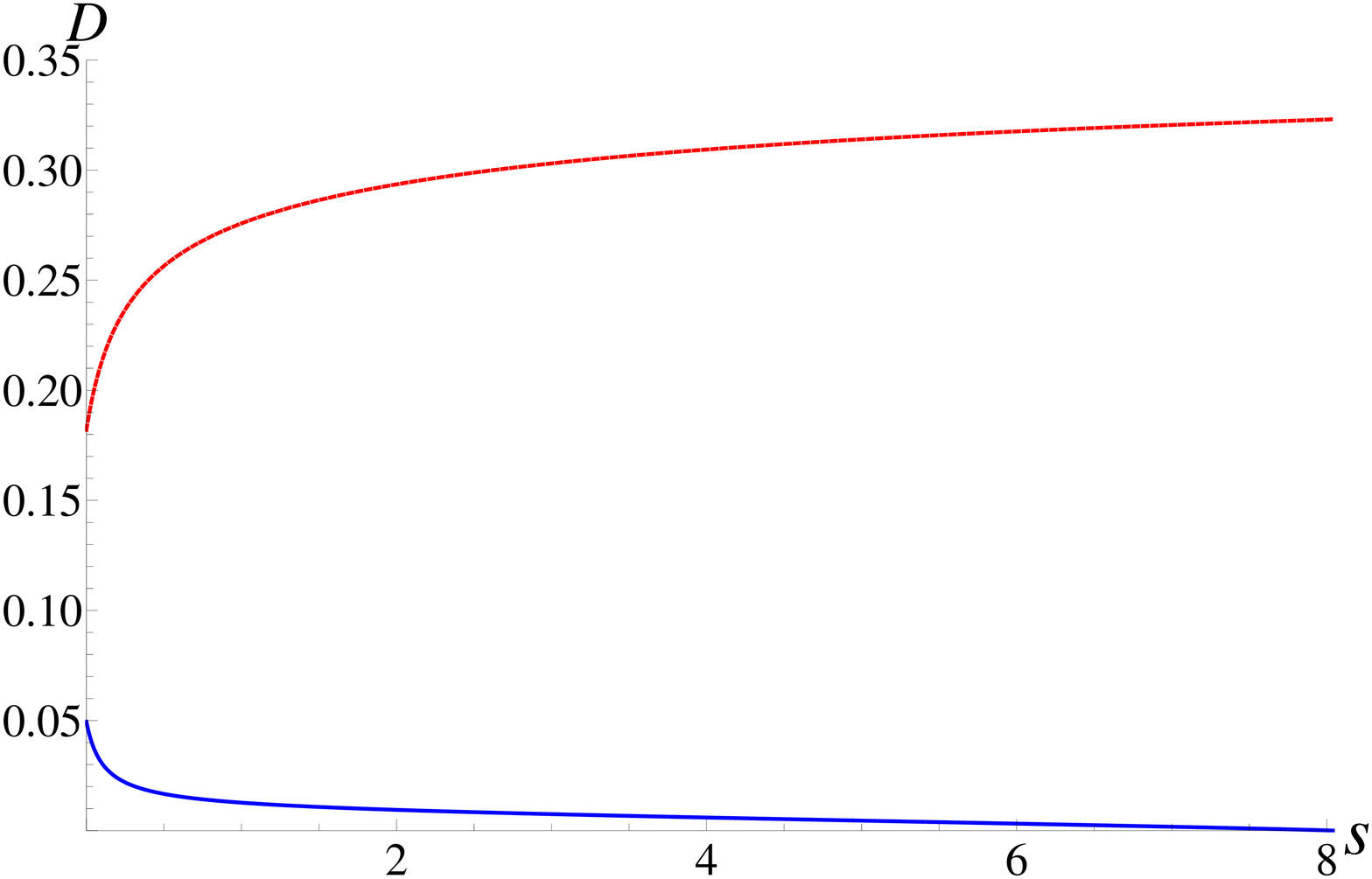}
\hskip 2.cm
\includegraphics[keepaspectratio, scale=0.21, angle=0]{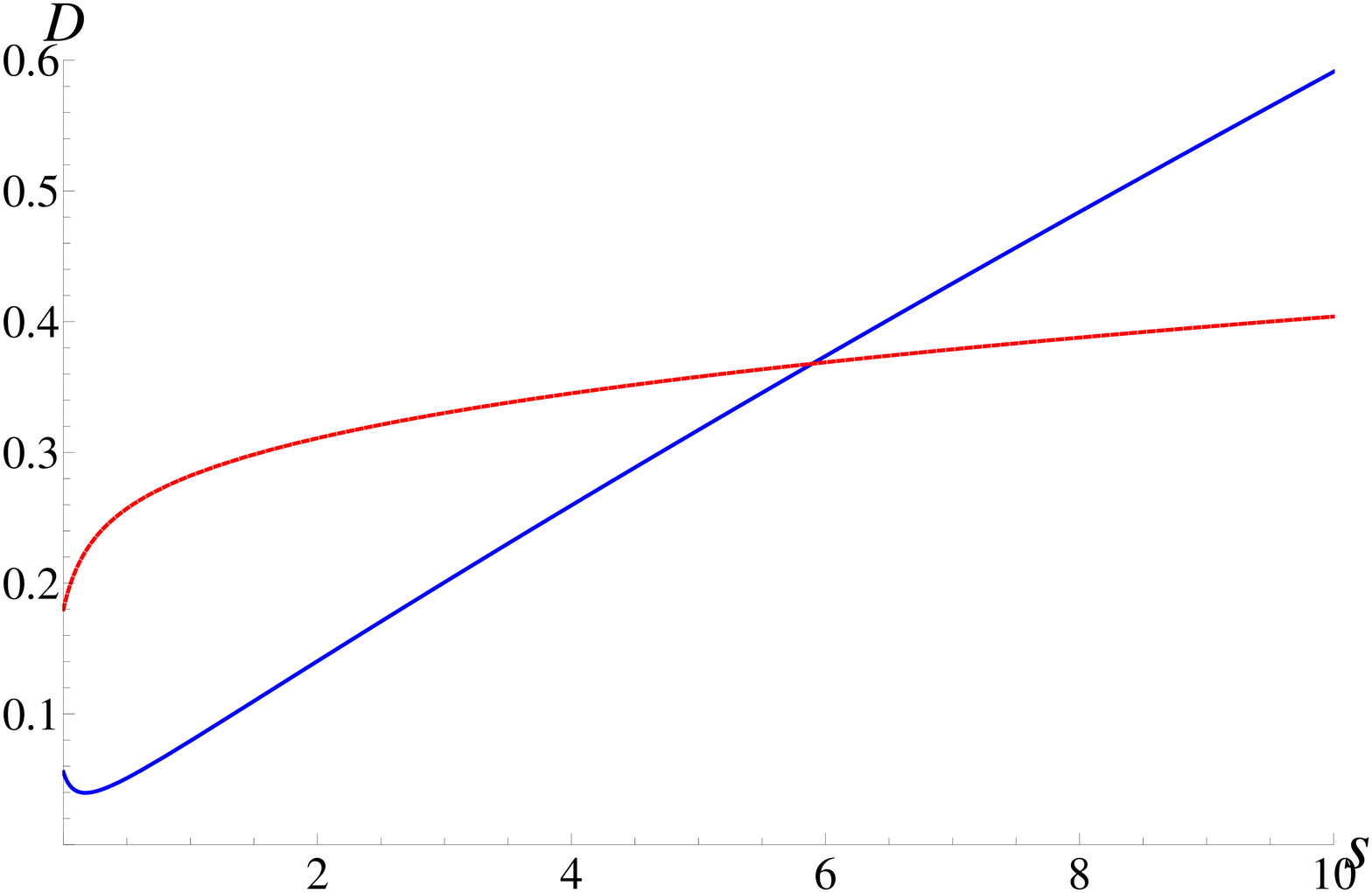}
\\
  \caption{\baselineskip 14pt 
 The radial null geodesic of 1-brane in the five-dimensional spacetime 
 (\ref{ge:bg:Eq}) is depicted. We have fixed $dt/ds|_{s=0}=-1$\,, 
 $c_0=-1$\,, $M=1$\,, in the solution (\ref{ge:bg:Eq}). 
 The red (dashed) and blue (solid) 
 lines correspond to the radial null geodesic $r/10$ 
 and function $h$, respectively. 
 "$D$" 
 in the vertical axis indicates the value of $r/10$, or 
 the function $h$, while "$s$" 
 in the horizontal axis denotes the affine parameter. 
 If we set $r(s=0)=1.82$ and $t(s=0)=-c_0^{-1}M [r(s=0)]^{-1}$, 
 the function $h$ settles down to zero 
 in left panel. The past directed null geodesic hits 
 the singularity. Then, the initially "smooth and 
 regular" data evolves into a 
 timelike singularity. 
 The right panel illustrates the case in which $r(s=0)$, $t(s=0)$ are
  given by $r(s=0)=1.8$, $t(s=0)=-c_0^{-1}M [r(s=0)]^{-1}$\,, respectively. 
  Since the null geodesic can extend to past null infinity, 
 it never actually reaches the singularity. 
 }
  \label{fig:p=1}
 \end{center}
\end{figure}

\begin{figure}[h]
 \begin{center}
\includegraphics[keepaspectratio, scale=0.21, angle=0]{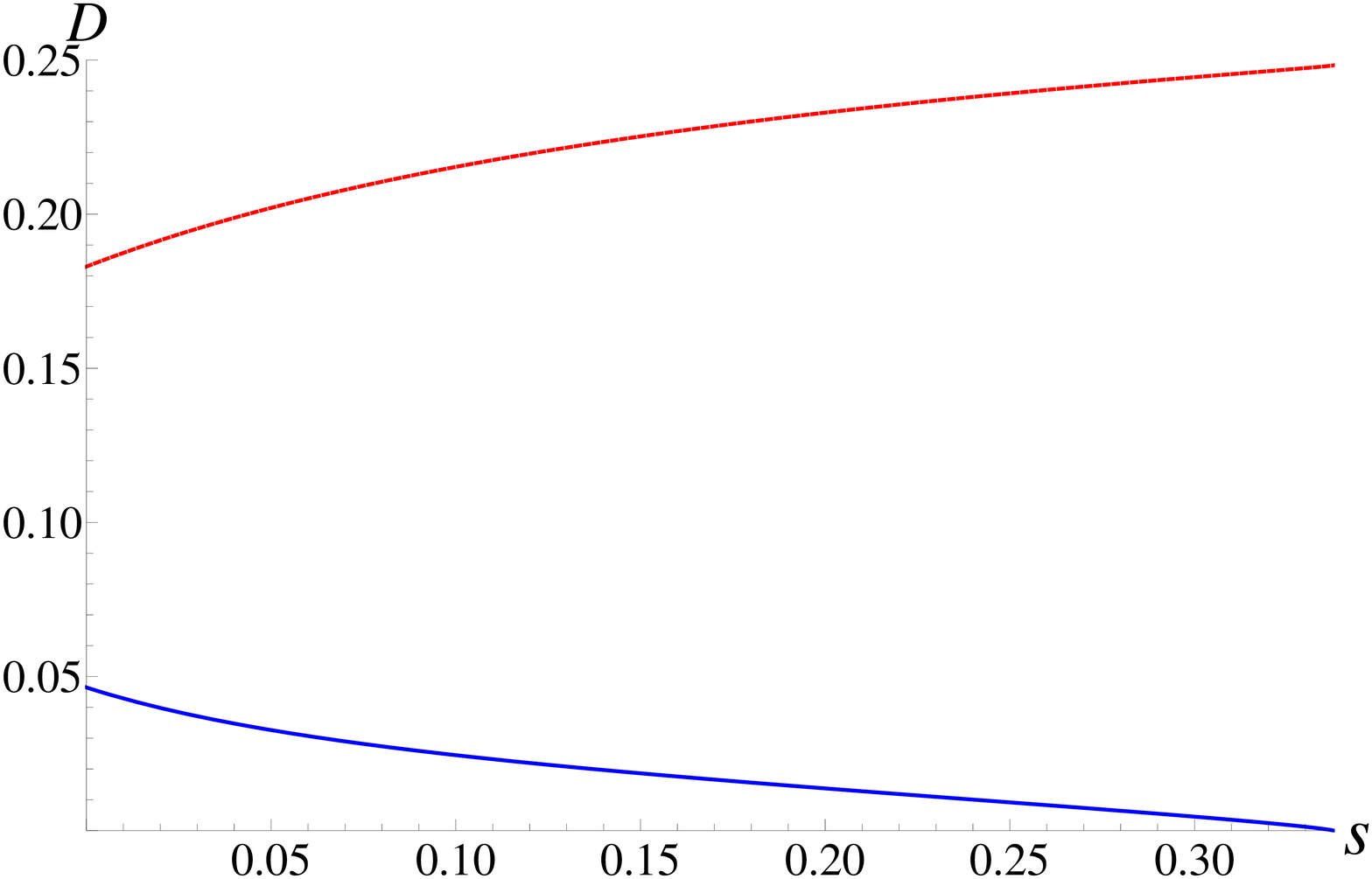}
\hskip 2.cm
\includegraphics[keepaspectratio, scale=0.21, angle=0]{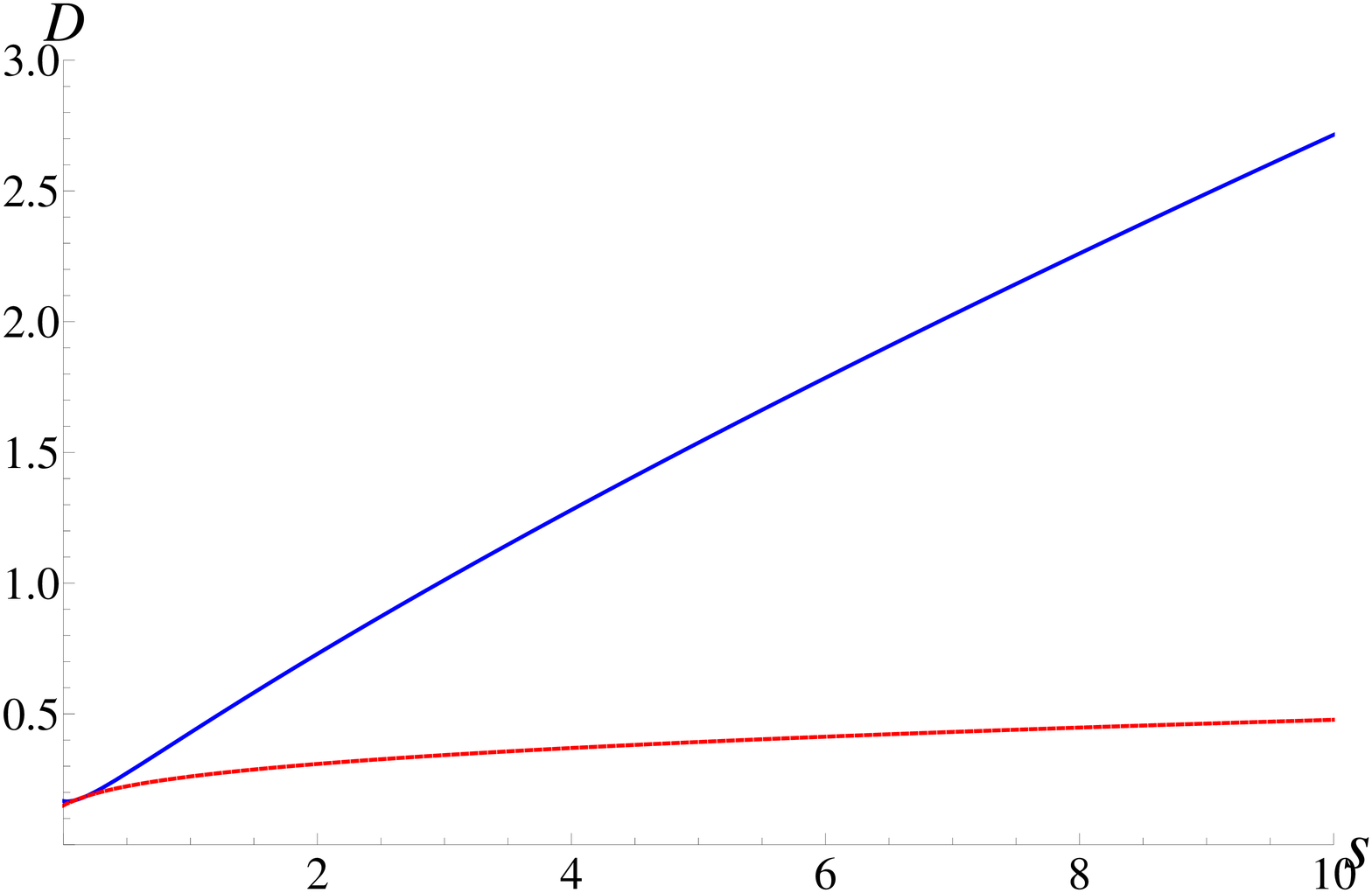}
\\
  \caption{\baselineskip 14pt 
 We give the radial null geodesic of 2-brane in the six-dimensional spacetime 
(\ref{ge:bg:Eq}). We have set $dt/ds|_{s=0}=-1$\,, $c_0=-1$\,, 
 $M=1$\,, in the solution (\ref{ge:bg:Eq}). 
 The red and blue curves show the behavior of radial null geodesic $r/10$ 
 and function $h$, 
 respectively.  "$D$" 
 in the vertical axis indicates the value of $r/10$, or 
 the function $h$, while "$s$" 
 in the horizontal axis denotes the affine parameter.
 The left panel illustrates the case in which the initial condition 
 of $r$, $t$ are  fixed by $r(s=0)=1.83$\,, 
 $t(s=0)=-c_0^{-1}M [r(s=0)]^{-1}$\,, respectively. 
 The geodesic hits a curvature   
 singularity in finite affine parameter, and the spacetime is null
  geodesically incomplete.  
  The "regular and smooth" initial data becomes 
  a timelike singularity as the time evolves. 
 In the right panel, we set $r(s=0)=1.5$\,, 
 $t(s=0)=-c_0^{-1}M [r(s=0)]^{-1}$\,. 
 The null geodesic reaches 
 past null infinity while it never hits the singularity. 
 }
  \label{fig:p=2}
 \end{center}
\end{figure}

\begin{figure}[h]
 \begin{center}
\includegraphics[keepaspectratio, scale=0.21, angle=0]{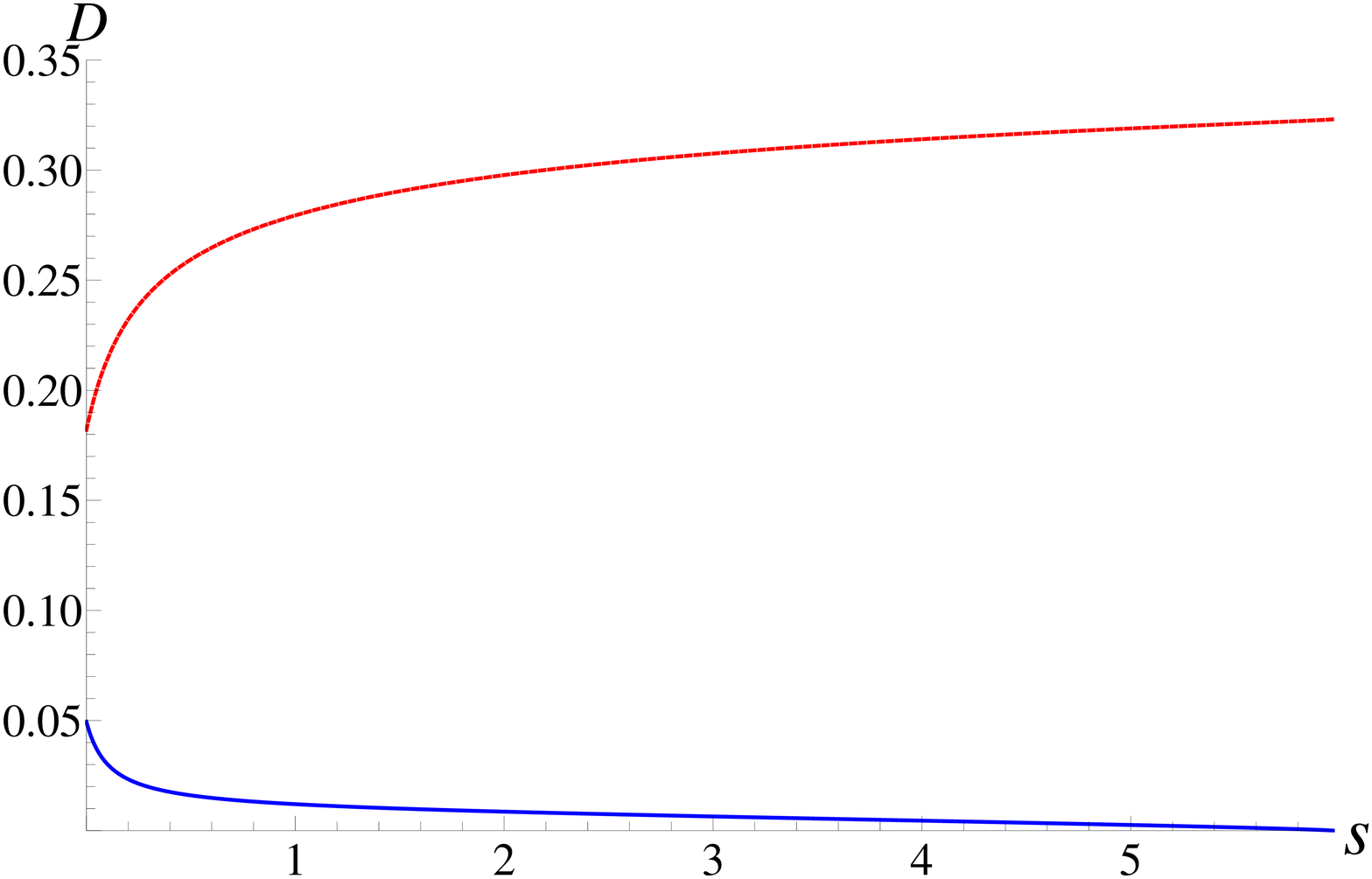}
\hskip 2.cm
\includegraphics[keepaspectratio, scale=0.21, angle=0]{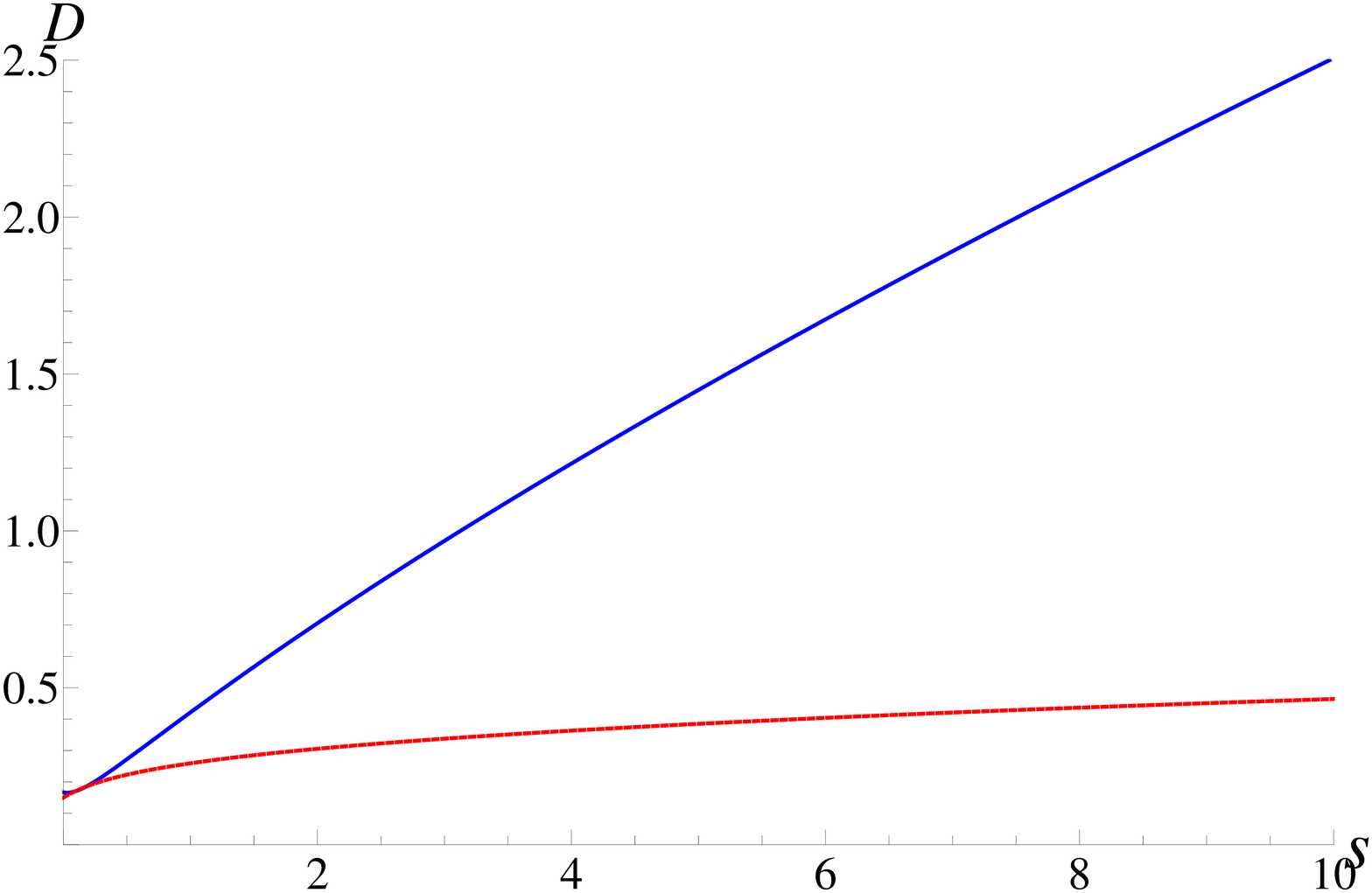}
\\
  \caption{\baselineskip 14pt 
 We illustrate the motion of radial null geodesic for 3-brane 
 in the seven-dimensional spacetime. 
In this plot, the parameters have been set to $dt/ds|_{s=0}=-1$\,,  
 $c_0=-1$\,, $M=1$\,, in the solution (\ref{ge:bg:Eq}). 
 We depict the dynamics of radial null geodesic $r/10$ (red line) 
 and function $h$ (blue line), respectively. 
 "$D$" 
 in the vertical axis indicates the value of $r/10$, or 
 the function $h$, while "$s$" 
 in the horizontal axis denotes 
 the affine parameter. 
 The left panel shows that the function $h$ approaches zero 
 for the present choice of parameters.  Here the initial data $r(s=0)$\,, 
 $t(s=0)$ have been fixed by $r(s=0)=1.82$\,, 
 $t(s=0)=-c_0^{-1}M [r(s=0)]^{-1}$\,, respectively. 
 The "nonsingular" initial conditions 
 evolve into a naked singularity. 
 A typical configuration 
 for fixing $r(s=0)=1.5$ 
 is also shown in right panel. The null geodesic never 
 reaches the timelike singularity. } 
  \label{fig:p=3}
 \end{center}
\end{figure}

\begin{figure}[h]
 \begin{center}
\includegraphics[keepaspectratio, scale=0.21, angle=0]{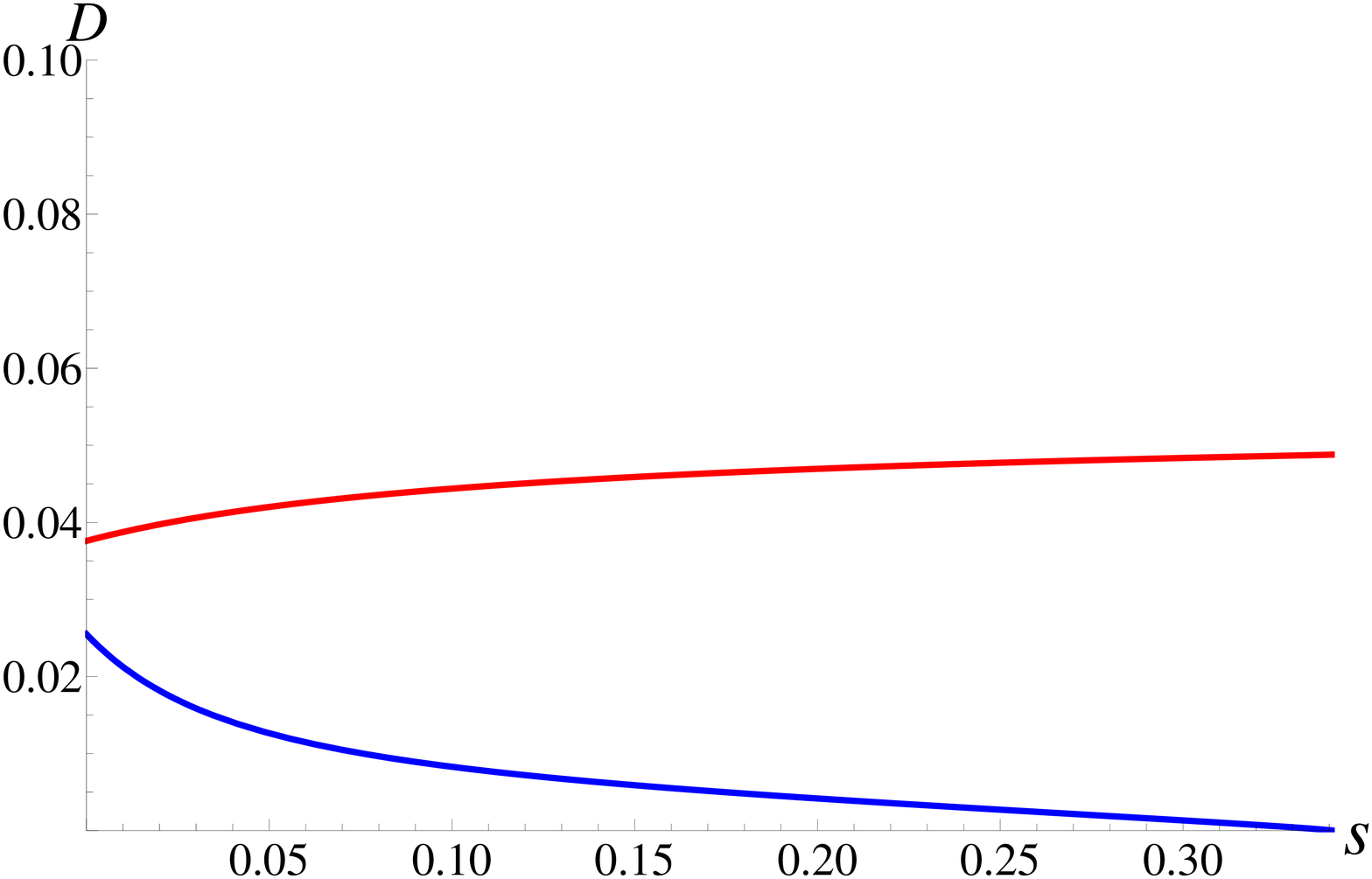}
\hskip 2.cm
\includegraphics[keepaspectratio, scale=0.21, angle=0]{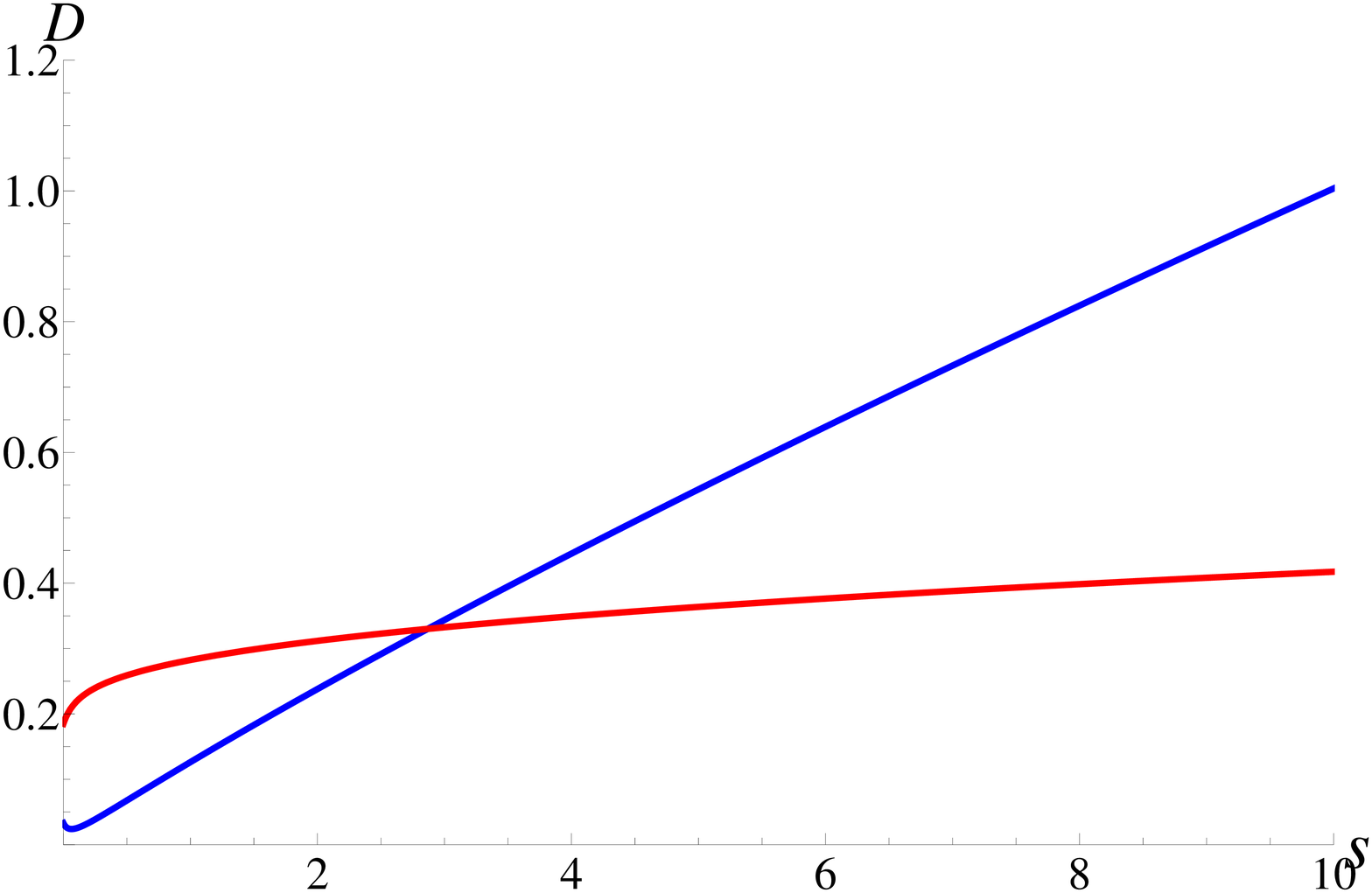} 
\\
  \caption{\baselineskip 14pt 
 We illustrate the motion of radial null geodesic for M5-brane 
 in the eleven-dimensional spacetime. 
 The parameters have been set to $dt/ds|_{s=0}=-1$\,,  
 $c_0=-1$\,,  $M=1$\,, in this plot. 
 In the left panel, "$D$" 
 in the vertical axis indicates the value of $r/50$ for red line, or 
 the function $h$ for blue line\,. 
 In the left panel, "$D$" 
 in the vertical 
 axis denotes the value of $r/10$ for red line, or 
 the function $h$ for blue line. 
 In both panels, "$s$"  
 in the horizontal axis shows the affine parameter. 
 The left panel shows that the function $h$ goes to zero 
 for the present choice of parameters.  The parameters $r(s=0)$ and 
 $t(s=0)$ have been set to $r(s=0)=1.88$\,, 
 $t(s=0)=-c_0^{-1}M [r(s=0)]^{-3}$\,, respectively\,. 
 One can note that the regular initial data 
 evolves into a naked singularity. Then, 
 this would allow one to find violations of cosmic censorship. 
 On the other hand, we have set $r(s=0)=1.85$\,, 
 $t(s=0)=-c_0^{-1}M [r(s=0)]^{-3}$\,, in right panel. 
 In this case, the null geodesic never hits 
 the timelike singularity. 
 }
  \label{fig:p=5}
 \end{center}
\end{figure}

Finally we calculate the radial null geodesics numerically 
in a dynamical $p$-brane background with $M\ne 0$ and $c_0=-1$\,. 
We show the motion of radial null geodesics for a 1-brane (Fig \ref{fig:p=1}), 
2-brane (Fig \ref{fig:p=2}), 3-brane (Fig \ref{fig:p=3}), 
and M5-brane (Fig \ref{fig:p=5}), respectively. 
The left panels (a) in Figs.~\ref{fig:p=1} - \ref{fig:p=5}   
show that past directed null geodesics reach the naked singularity 
in finite affine parameter.  
The "nonsingular" initial data evolves into a 
timelike singularity. Since we can set a "regular and smooth" initial data 
for the M5-brane, this leads to be a violation of cosmic censorship. 
On the other hand, 
we illustrate that past directed null geodesics 
can reach past null infinity 
in the right panels (b) in Figs.~\ref{fig:p=1} - \ref{fig:p=5}\,. 

\begin{figure}[h]
 \begin{center}
%
\includegraphics[keepaspectratio, scale=0.31, angle=-90]{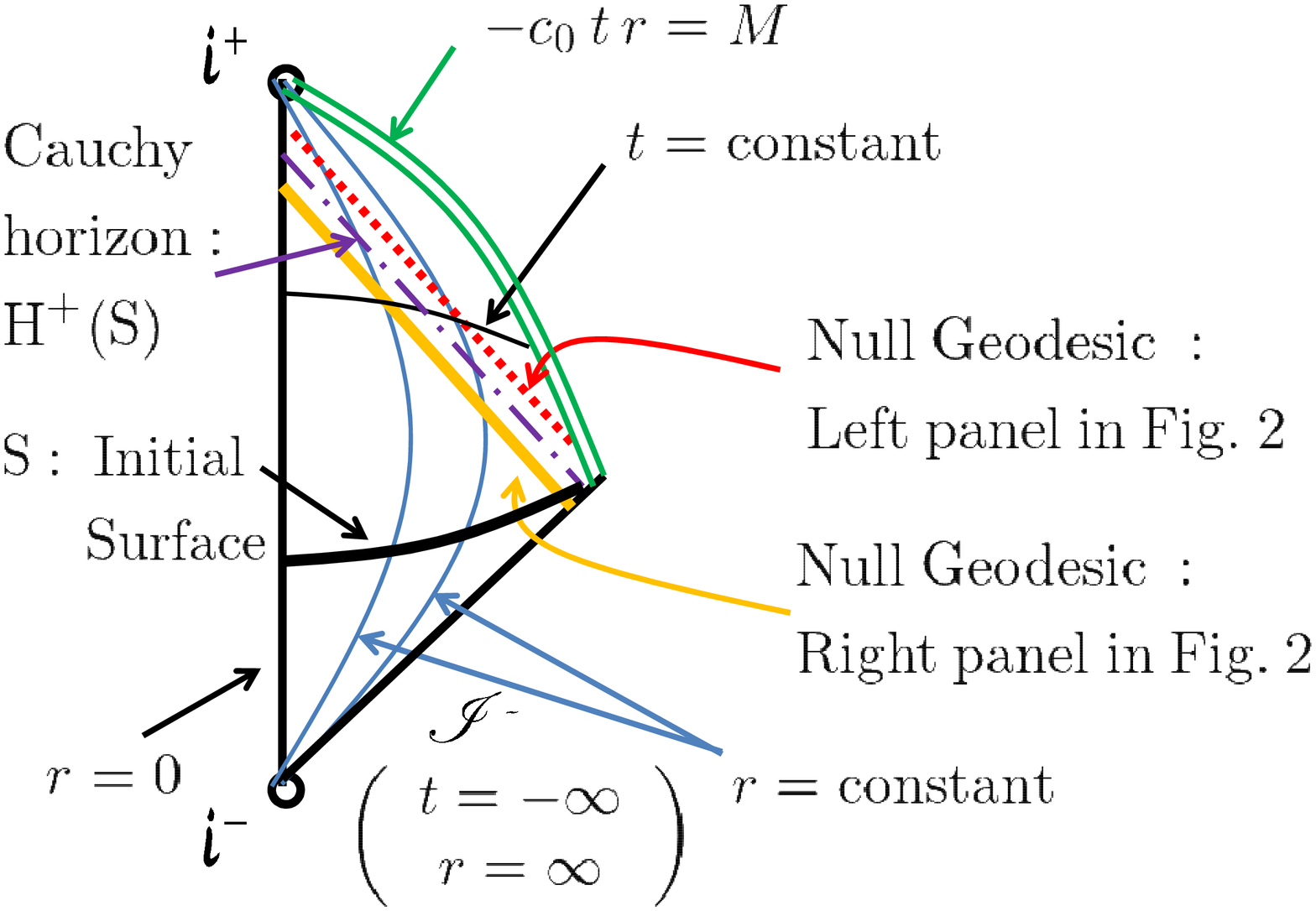}
\hskip 0.5cm
\includegraphics[keepaspectratio, scale=0.31, angle=-90]{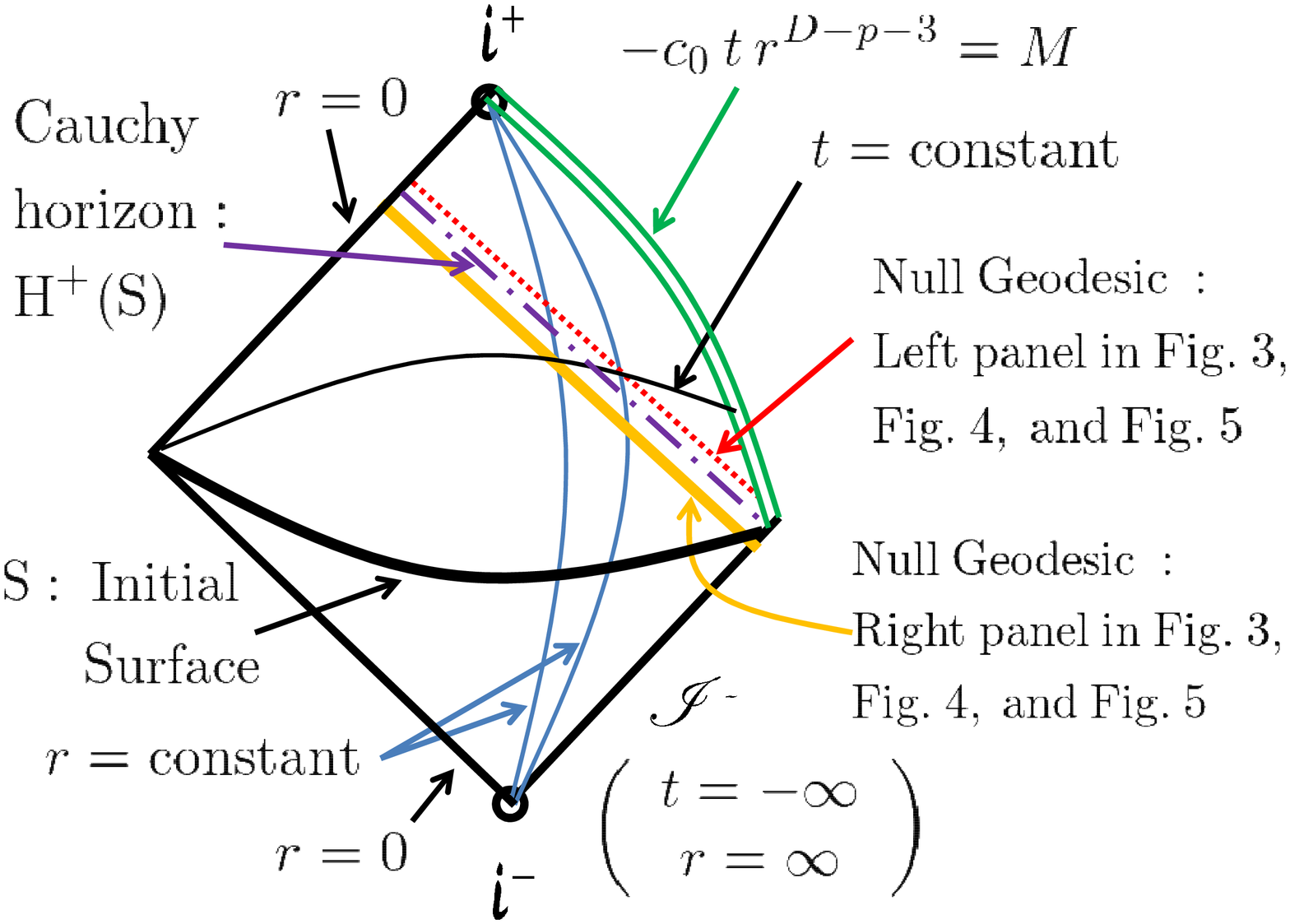}
\\
  \caption{\baselineskip 14pt 
The Penrose diagram of the $D$-dimensional spacetime 
(\ref{ge:bg:Eq}) with $M\ne 0$\,, $c_0<0$\,, 
 $D-p-3=1$ 
 is depicted in left panel. 
The right panel in this figure shows Penrose diagram  
for the solution (\ref{ge:bg:Eq}) with $c_0<0$\,, 
$D-p-3\ge 2$\,. 
In both cases, the green line (double line) at 
$h=0$ is the timelike singularity. The dot-dash (purple) line shows the 
Cauchy horizon H${}^+$(S).  
In the domain $D^+$(S) which is inside of a Cauchy horizon H${}^+$(S), the 
null geodesics passing through the initial surface S  
evolves far into the past infinitely. 
We focus on the behavior of null geodesic near 
the double line. 
The naked singularity appears first at large $r$ as the time increases.  
The past directed null geodesics (red line or dotted line) 
corresponding to the left panel in Figs.~\ref{fig:p=1}, \ref{fig:p=2}, 
\ref{fig:p=3}, and \ref{fig:p=5} 
reach a singularity in finite affine parameter while another 
past directed null geodesics (orange line or bold line) corresponding to 
the right panel in Figs.~\ref{fig:p=1}, \ref{fig:p=2},  \ref{fig:p=3}, and 
\ref{fig:p=5} never hit the
singularity. The difference between them comes from 
the initial data.  
We see that the asymptotic behavior of the null curves 
depends crucially on whether $r$ is inside or outside the  
Cauchy horizon. 
If the null curve is outside of the Cauchy horizon, 
the past directed radial null geodesics always reach the singularity. 
The dotted (red) line shows that the  
"regular and smooth" initial data  
evolves into a timelike curvature singularity. For M-brane, we can set a 
perfectly smooth and regular initial data. Then, the null geodesics 
in the dynamical M-brane background  
hit a timelike curvature singularity at $h=0$\,. 
This suggests a violation of cosmic censorship 
in the dynamical $p$-brane background. 
}
  \label{fig:tp}
 \end{center}
\end{figure}

The Penrose diagram for the entire spacetime with $M\ne 0$, 
is shown in Fig. \ref{fig:tp}. 
We depict two past directed null geodesics in Fig. \ref{fig:tp}. 
Although the red (dotted) curve hits the naked singularity in finite
affine parameter, the orange (bold) line  
can avoid the singularity. 
One can note that there is a null hypersurface $r=r_{\rm h}(t)$  
in the dynamical $p$-brane background. 
The null geodesics outside of it hit the timelike singularity while 
the past directed radial null geodesics inside of this hypersurface 
can reach past null infinity. 
There is a Cauchy horizon (dot-dash line) in the dynamical $p$-brane solution 
(\ref{ge:bg:Eq}). 
In the domain $D^+$(S) which is inside of a Cauchy horizon H${}^+$(S), the 
null geodesics can pass through the initial surface S and  
evolves far into the past infinitely.

We will see that the $p$-brane background changes the causal structure of 
the dynamical solution without $p$-brane. 
When one adds a single $p$-brane, 
the spacelike singularity is replaced by a timelike one which is visible 
to observers in the spacetime. So this solution has a 
naked singularity. 
The naked singularity appears first at large $r$, and 
the null geodesic cannot be extended beyond 
$r=\left(-M/c_0\,t\right)^{1/D-p-3}$\,. 

The dynamical $p$-brane backgrounds in general have a curvature 
singularity at $r=0$\,, if the dilaton is non-trivial. 
The near $p$-brane geometry is the same as the static one 
because $h(t, r)\rightarrow h_1(r)$ as $r\rightarrow 0$\,. 
Then the geometry approaches the static solution. Although we expect that 
the singularity at $r=0$ can be resolved in string theory
\cite{Johnson:1999qt, Jarv:2000zv, Peet:1999nh, Johnson:2000ch, 
Yamaguchi:2001yd}, the timelike singularity at $h=0$ is still left.  
If the dynamical $p$-brane background has a horizon geometry, 
we can regard the present 
time-dependent solution as a black hole. In fact, we know
that M-branes give 
black hole spacetimes
in the static limit due to vanishing dilaton \cite{Maeda:2009zi}. 
From the Eq.~(\ref{n:bM:Eq}), the dynamical M-brane background becomes 
regular at $y^i=y^i_l$ 
because of the Ricci scalar for the M-brane being constant. 
Since the perfectly smooth and regular initial data in the far past 
evolves into a timelike curvature singularity at $h=0$, 
the cosmic censorship is violated in the M5-brane background. 

\subsection{The dynamical $p$-brane depending on the null coordinate}
\label{sec:null}
Here we discuss the radial null geodesics on 
the dynamical $p$-brane background whose metric are given by 
the null coordinates \cite{Binetruy:2007tu, Maeda:2009tq}:
\Eqrsubl{n:bg:Eq}{
&&ds^2=h^a(u, r)\left(-2du\,dv+\delta_{PQ}dx^Pdx^Q\right)
+h^b(u, r)\left(dr^2+r^2w_{mn}dz^mdz^n\right)\,,
 \label{n:metric:Eq}\\
&&u=\frac{1}{\sqrt{2}}\left(t-x\right)\,,~~~~~~
v=\frac{1}{\sqrt{2}}\left(t+x\right)\,,\\
&&h(u, r)=h_0(u)+h_1(r),~~~~~h_0(u)=c_0 u
\,,~~~~~
h_1(r)=\frac{M}{r^{D-p-3}}\,, 
  \label{n:h:Eq}
}
where $\delta_{PQ}$ is the metric of $(p-1)$-dimensional Euclid space, 
and we assume $d_\Ysp=0$\,. 
Since the geodesic is null, the geodesic equation yields 
\Eqrsubl{n:null:Eq}{
&&\dot{u}=fh^{-a}\,,~~~~\dot{v}=f^{-1}h^b\dot{r}^2\,,\\
&&\ddot{r}=\frac{(a+b)(D-p-3)M}{2h}\frac{\dot{r}^2}{r^{D-p-2}}
-b\,f\,c_0\,h^{-b}\,\dot{r}\,,
}
where $f$ is constant. 
First we study the null geodesic without $p$-brane. 
The solution of geodesic equation with $M=0$ is given by 
\Eqrsubl{n:geodesic:Eq}{
u(s)&=&\left[(1+a)\left(c_0^{-a}\,f\,s+u_0\right)\right]
^{\frac{1}{1+a}},\\
r(s)&=&r_0\ln\left|c_0^{-a}\,f\,s+u_0\right|+r_1\,,
}
where $u_0$, $r_0$, $r_1$ are constants\,. 
If we set $r_0<0$\,, 
the singularity in this background (\ref{n:bg:Eq}) 
 is not spacelike but null, so that as $u\rightarrow 0$, 
 $r\rightarrow \infty$\,.
Hence, observers do not lose causal contact 
as they approach the singularity. This is an important difference 
between the solution with $M=0$\,, 
and standard FLRW 
universe with spacelike singularity.

\begin{figure}[h]
 \begin{center}
\includegraphics[keepaspectratio, scale=0.37, angle=0]{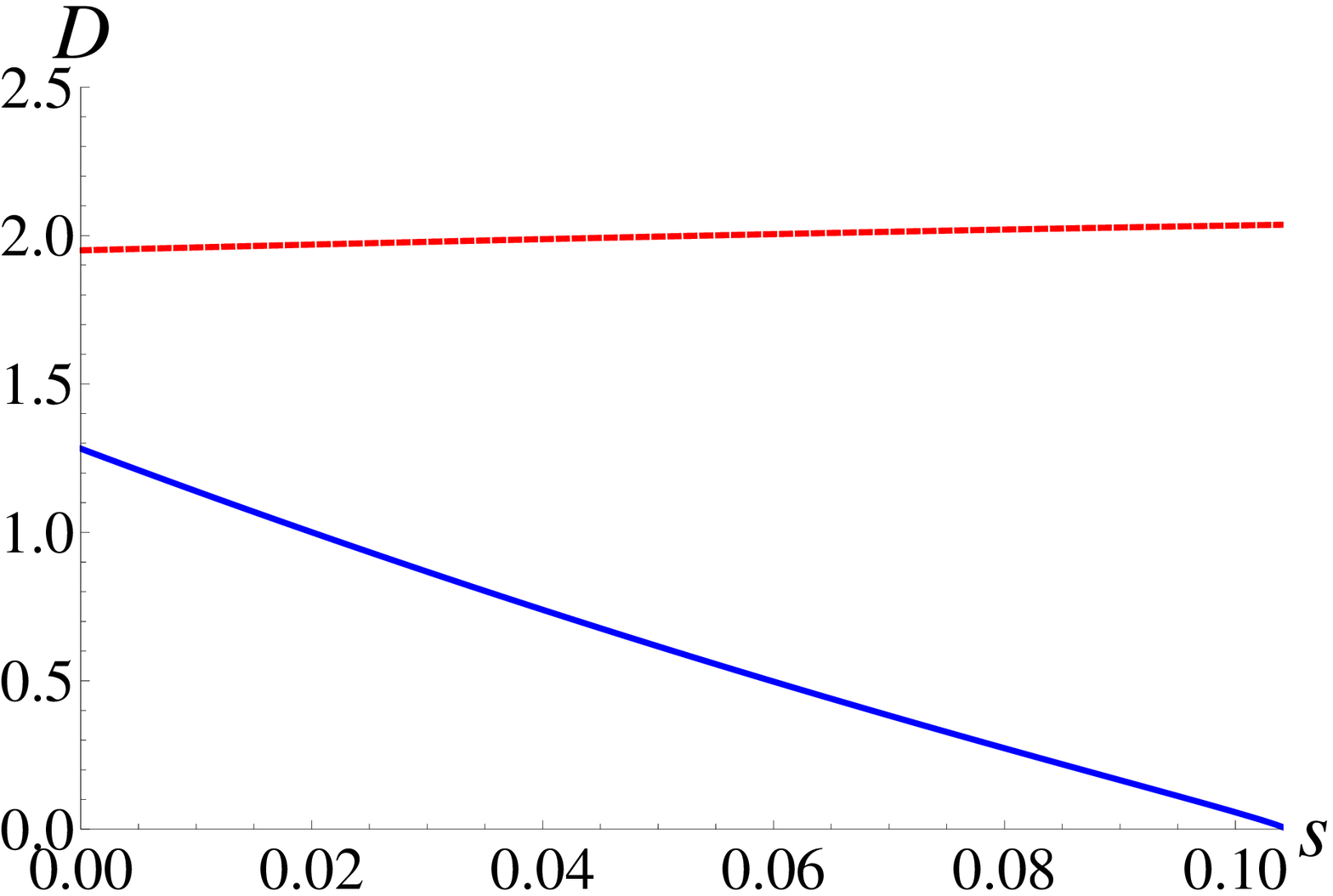}
\hskip 2.cm
\includegraphics[keepaspectratio, scale=0.37, angle=0]{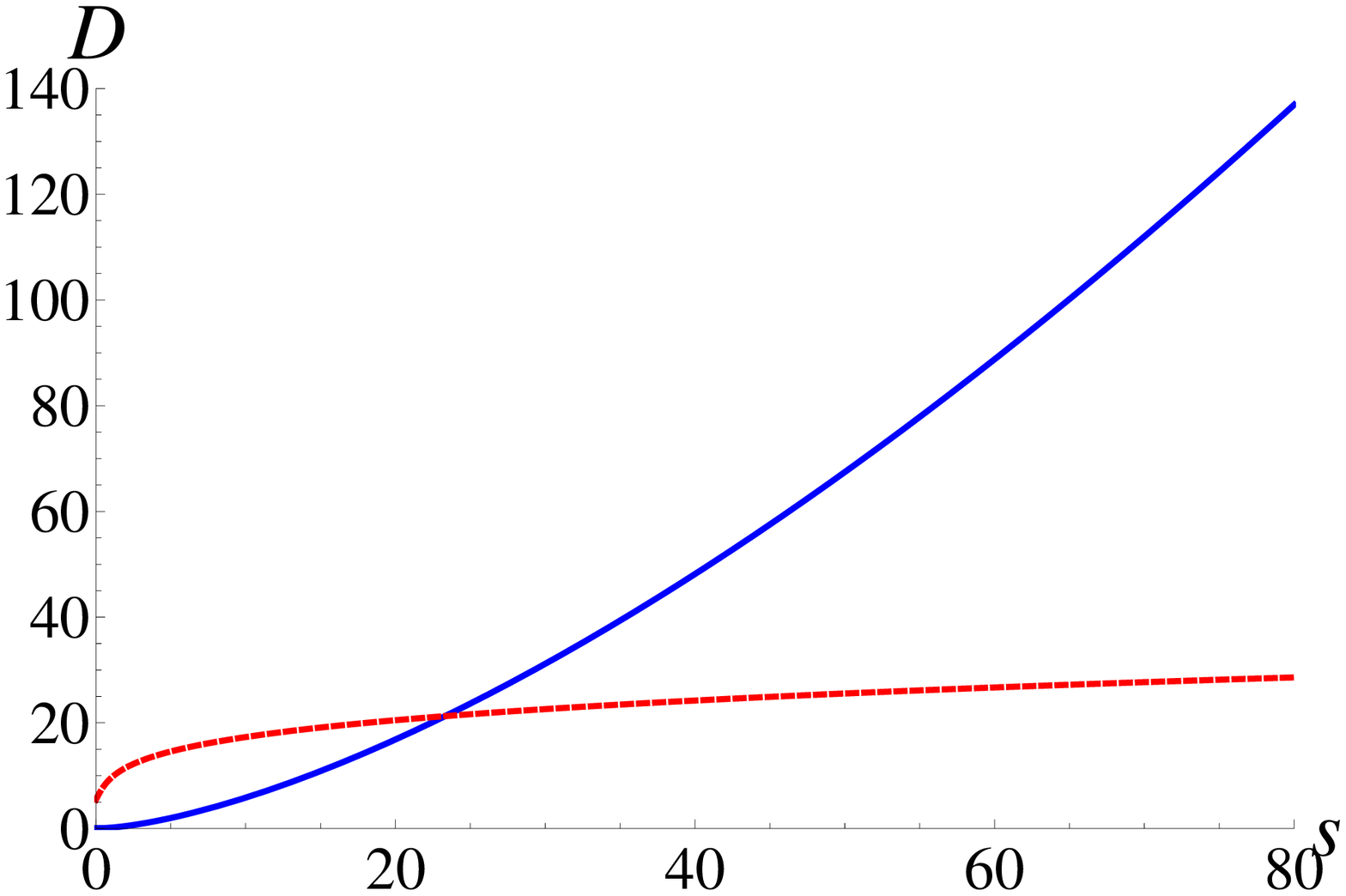}
\\
  \caption{\baselineskip 14pt 
 The radial null geodesic of 1-brane in the five-dimensional spacetime 
 (\ref{n:metric:Eq}) is depicted. We have fixed $dr/ds|_{s=0}=1$\,, 
 $c_0=-1$\,, $f=-1/2$\,,  
 $M=1$\,, in the solution (\ref{n:bg:Eq}). 
 The red (dashed) and blue (solid) lines correspond to the radial 
 null geodesic $r$ 
 and function $h$, respectively. 
 In the left panel, "$D$"  
 in the vertical axis indicates the value of $r$ for red line, or 
 the function $10^2\,h$ for blue line\,. 
 On the other hand, "$D$" 
 in the vertical 
 axis gives the value of $\e^r$ for red line, or 
 the function $h$ for blue line in the right panel. 
 In both panels, "$s$"  
 in the horizontal axis denotes the affine parameter. 
 If we set the initial value of radial coordinate and time 
 $r(s=0)=1.95$\,, $u(s=0)=-c_0^{-1}M [r(s=0)]^{-1}$\,, 
 a singularity appears at $s=0.104$ in left panel.
 The past directed null geodesic hits 
 the timelike singularity. 
 The "smooth and regular" initial data evolves 
 into a timelike singularity. 
 The right panel shows the case in which $r(s=0)$\,, $u(s=0)$ are 
 given by $r(s=0)=1.7$\,, $u(s=0)=-c_0^{-1}M [r(s=0)]^{-1}$\,, respectively. 
 The null geodesic can reach past null infinity. 
  }
  \label{fig:np1}
 \end{center}
\end{figure}

\begin{figure}[h]
 \begin{center}
\includegraphics[keepaspectratio, scale=0.37, angle=0]{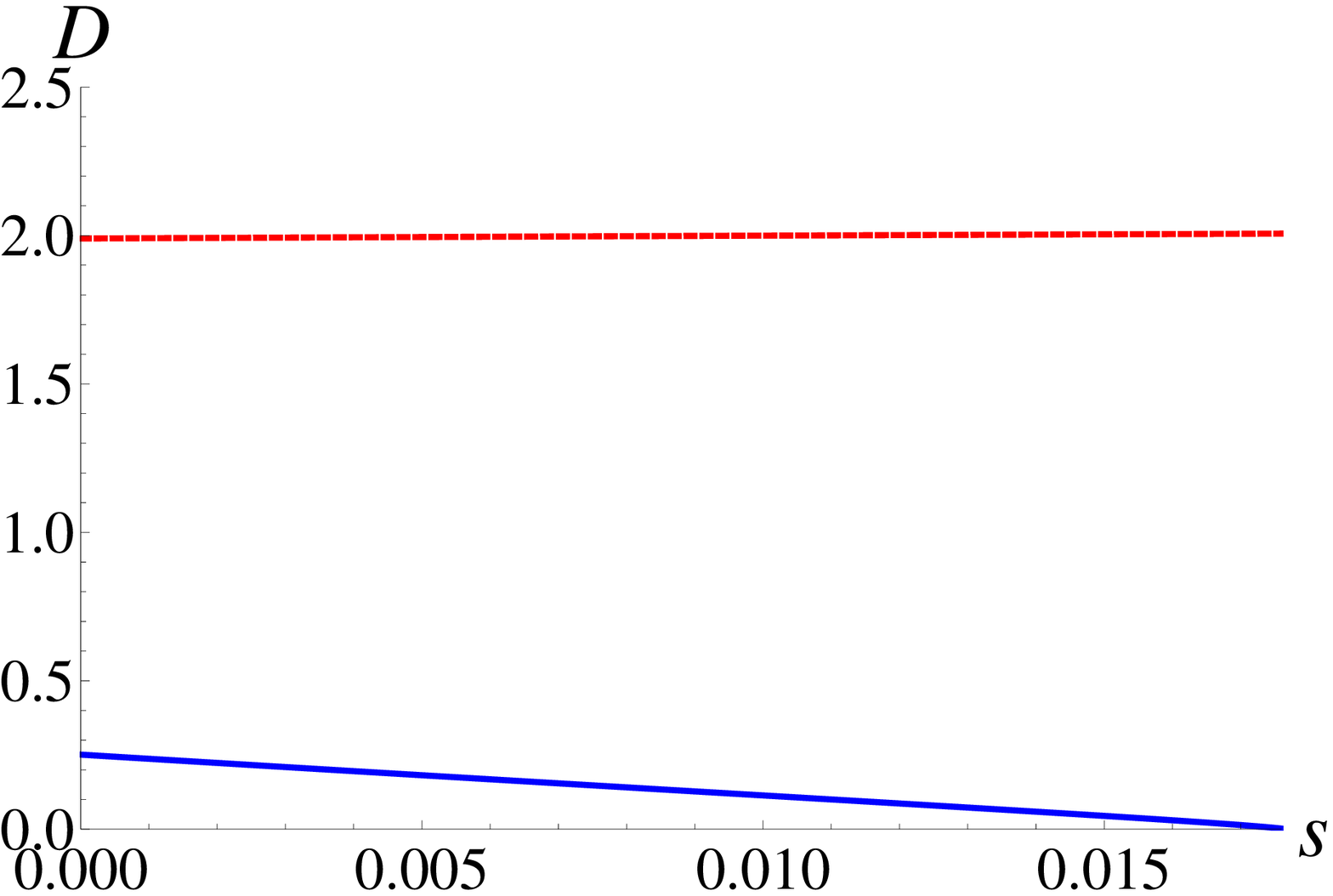}
\hskip 2.cm
\includegraphics[keepaspectratio, scale=0.37, angle=0]{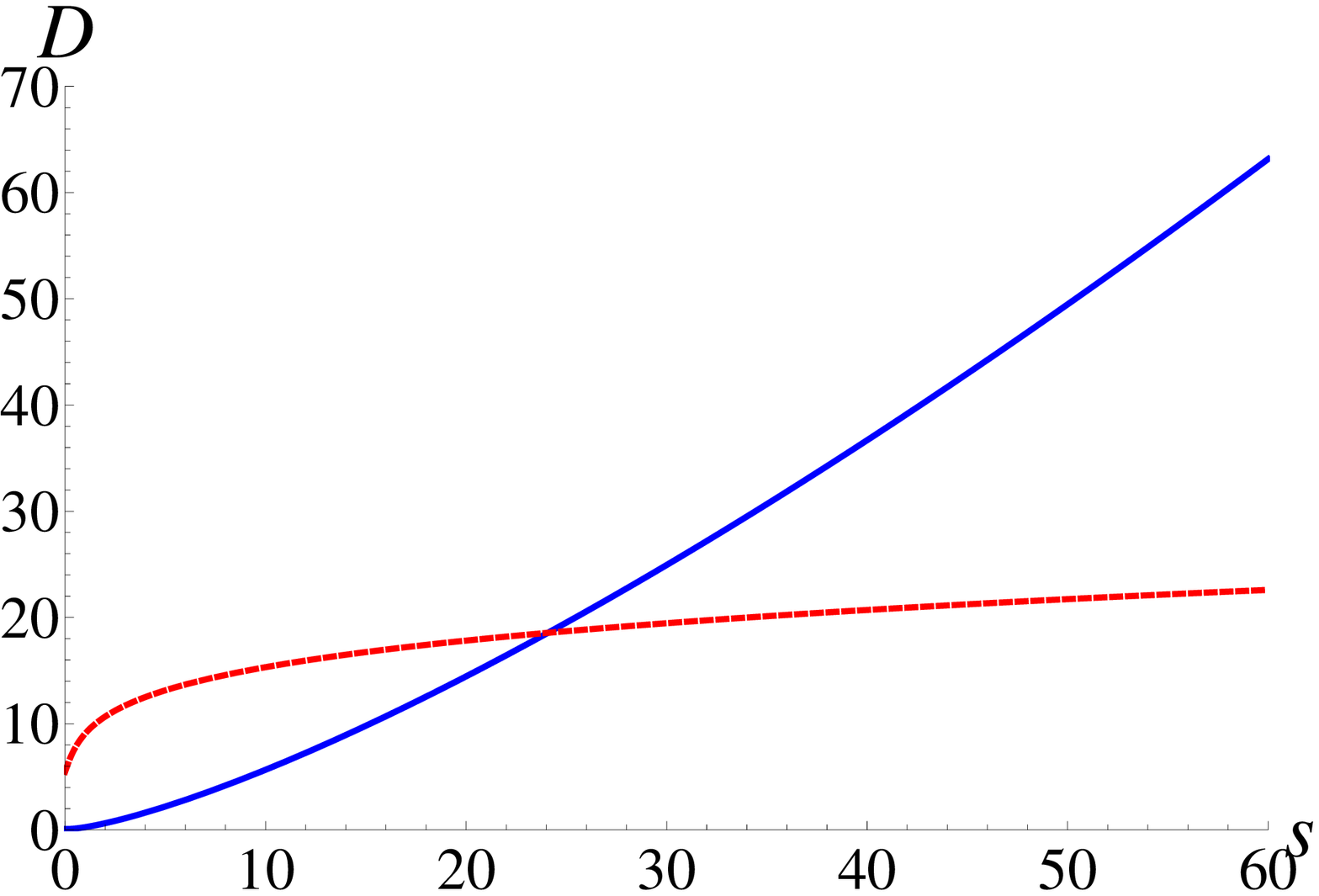}
\\
  \caption{\baselineskip 14pt 
 We give the radial null geodesic of 2-brane in the six-dimensional spacetime 
 (\ref{n:metric:Eq}). 
 We have fixed $dr/ds|_{s=0}=1$\,, 
 $c_0=-1$\,,  $f=-1/2$\,, 
 $M=1$\,, in the solution (\ref{n:bg:Eq}). 
 The red (dashed) and blue (solid) lines correspond to the 
 past directed radial null geodesic $r$ 
 and function $h$, respectively. "$D$"  
 in the vertical axis shows the value of $r$ for red line, or 
 the function $10^2\,h$ for blue line in the left panel. 
 In the right panel, 
 "$D$" in the vertical 
 axis indicates the value of $\e^r$ for red line, or 
 the function $h$ for blue line. In both panels, "$s$" 
 in the horizontal axis denotes the affine parameter. 
 The left panel shows the case in which the initial condition 
 of $r$\,, $u$ are fixed by $r(s=0)=1.99$\,, 
 $u(s=0)=-c_0^{-1}M [r(s=0)]^{-1}$\,, respectively. 
 Since the function $h$ eventually vanishes at $s=0.0176$\,. 
 the past directed radial null geodesic hits a timelike   
 singularity. The "regular and smooth" initial data evolves 
 into a timelike curvature  singularity. 
 The right panel depicts the past directed radial null geodesic 
 which does not reach the singularity. The initial value of $r$\,, $u$ are 
 fixed by $r(s=0)=1.70$\,, $u(s=0)=-c_0^{-1}M [r(s=0)]^{-1}$\,, respectively. 
 }
  \label{fig:np2}
 \end{center}
\end{figure}

Next we solve the radial null geodesic equation numerically 
in the dynamical $p$-brane background (\ref{n:bg:Eq}) 
in the case of $M\ne 0$ and $f\ne 0$\,. 
The motion of radial null geodesics for 1-brane, 2-brane   
are shown in Figs.~\ref{fig:np1} and \ref{fig:np2}, respectively. 
The left panels (a) in Figs.~7 and 8 
illustrate the past directed null geodesics reach the timelike singularity 
in finite affine parameter.  
We show that past directed null geodesics can reach past null infinity
in the right panels (b) in Figs.~\ref{fig:np1} and \ref{fig:np2}. 
Although the evolution of the null geodesic depends on the initial data, 
the "regular and smooth" initial data 
in the far past (\ref{n:bg:Eq}) 
evolves into a spacetime with a naked singularity.

\section{Discussions}
  \label{sec:discussions}
In the present paper, we have discussed the causal structure and the 
behavior of the singularity in the dynamical $p$-brane background 
which can be viewed as a family of time dependent solutions 
to the Einstein-Maxwell-dilaton theory \cite{Horne:1993sy}. 
This background gives an example to test cosmic censorship. 
The dynamical solution in this paper describes an arbitrary number 
of $p$-branes or 
extremal charged black holes in a expanding or contracting 
universe \cite{Maeda:2009zi}. 
The spacetime collapses to a spacelike or null singularity 
without $p$-branes. 
However, the spacelike or null singularity turns into a naked one  
if we add a single $p$-brane. 
Hence, regular initial data in the far past 
evolves into a timelike curvature singularity  
even if there is only one $p$-brane. 
This implies that cosmic censorship fails 
for the dynamical $p$-brane background. 
These results are consistent with the analysis of Horne \& Horowitz 
\cite{Horne:1993sy}. 

As we have said in section \ref{sec:gp}, the dynamical $p$-brane backgrounds 
 with non-trivial dilaton have singularities   
at $r=0$ and $h=0$\,. 
In the $p$-brane solutions, 
the event horizon shrinks 
down to zero size in the extremal limit. 
Then there is singularity at $r=0$ in the $p$-brane backgrounds. 
These solutions describe an extremal black hole in the expanding or 
collapsing universe \cite{Maeda:2009zi, Horne:1993sy}. 
The singularities at $h=0$ in the solutions 
(\ref{ge:bg:Eq}) and (\ref{n:bg:Eq}) are
 generated by the gravitational collapse. 
If we consider the $p$-brane solution which moves slightly away from 
the extremal limit, the spacelike surface $r=0$ will become a 
nonsingular horizon shielding a spacelike or null singularity inside. 
The slight change of the brane charge does not modify
the solution asymptotically. 
Since the timelike singularity at large $r$ will be present, 
nonsingular initial data evolves into a naked singularity.  
This implies that cosmic censorship is violated in 
the non extremal $p$-brane background. 
However, in order to study whether cosmic censorship fails
for this background, we have to construct the non-extremal dynamical 
$p$-brane solution. 

If the background has the vanishing or trivial dilaton, 
there is dynamical brane solutions whose have event horizon at $r=0$\,. 
Then, we can 
set smooth initial data evolving 
into a timelike curvature singularity. 
In this paper, we present that the cosmic censorship is violated 
for the dynamical M5-brane system in eleven-dimensional supergravity.

The cosmic censorship 
\cite{Penrose:1964wq, Penrose:1969pc}, which was supposed long times ago 
following the recognition of the special 
role of forming the singularity, 
is violated by the time-dependence of fields in the context of general 
relativity. However, in the region of the dynamical $p$-brane 
near the naked  
singularity, the curvature of the spacetime becomes very large. 
Then we have to indeed 
discuss this issue in the quantum theory of gravity, 
such as string theory. Moreover, the string loop correction 
will be important if observers approach the singularity on the basis of string 
theory.  Hence, we can say very little at this time about the violation of 
the cosmic censorship in string theory unless higher order string $\alpha'$ 
corrections in the solution is included. 

\section*{Acknowledgments}
We thank Akihiro Ishibashi for valuable discussions and careful 
reading of the manuscript and also thank Kei-ichi Maeda, Norihiro Iizuka for 
discussions and valuable comments. This work was supported in
part by JSPS KAKENHI Grant Number 23740200 (KM).

\section*{Appendix}
\appendix

\section{Dynamical $p$-brane background}
\label{sec:ap} 
In this appendix, we work out the explicit form of 
the components of Einstein equations for a general metric 
which is compatible with the expected form of the sought for 
dynamical $p$-brane solutions.

For the $D$-dimensional action (\ref{p:a:Eq}), 
the field equations are given by
\Eqrsubl{ap:equations:Eq}{
&&\hspace{-0.5cm}R_{MN}=
\frac{1}{2}\pd_M\phi \pd_N \phi\nn\\
&&~~~~~~+\frac{1}{2}\frac{\e^{\epsilon c \phi}}
{\left(p+2\right)!}
\left[\left(p+2\right)
F_{MA_2\cdots A_{\left(p+2\right)}} 
{F_N}^{A_2\cdots A_{\left(p+2\right)}}
-\frac{p+1}{D-2}\,g_{MN}\,F^2_{\left(p+2\right)}\right],
   \label{ap:Einstein:Eq}\\
&&\hspace{-0.5cm}\lap\phi-\frac{1}{2}\frac{\epsilon c}
{\left(p+2\right)!}
\e^{\epsilon c\phi}F^2_{\left(p+2\right)}=0\,, 
   \label{ap:scalar:Eq}\\
&&\hspace{-0.5cm}
  d\left[\e^{\epsilon c \phi}\ast F_{\left(p+2\right)}\right]=0\,,
   \label{ap:gauge:Eq}
}
where $R_{MN}$, $\lap$ denote the Ricci tensor, Laplace operator 
with respect to the $D$-dimensional metric $g_{MN}$\,, respectively and 
the constants $c$\,, $\epsilon$ are defined by 
\Eqrsubl{ap:parameters:Eq}{
c^2&=&N-\frac{2(p+1)(D-p-3)}{D-2},
   \label{ap:c:Eq}\\
\epsilon&=&\left\{
\begin{array}{cc}
 +&{\rm if~the}~p-{\rm brane~is~electric}\\
 -&~~~{\rm if~the}~p-{\rm brane~is~magnetic}\,.
\end{array} \right.
 \label{ap:epsilon:Eq}
   }
Here $N$ is constant. 
We will consider general metrics with components depending on not only 
coordinates of transverse space to $p$-brane but also worldvolume ones. 
Hence, the time is not translational invariant. To begin with
we do not impose any symmetry in $D$-dimensional spacetime. 
Then we consider a particular case in which we have indeed 
$(p+1)$-dimensional Minkowski spacetime and $(D-p-1)$-dimensional 
Euclidean space. 
The $D$-dimensional metric of dynamical $p$-brane is written in the 
general form 
\Eqr{
&&\hspace{-0.7cm}
ds^2=h^a(x, y)q_{\mu\nu}(\Xsp)dx^{\mu}dx^{\nu}+h^b(x, y)\gamma_{ij}
(\Ysp)dy^idy^j\,,
 \label{ap:metric:Eq}
}
where $q_{\mu\nu}(\Xsp)$ is the $(p+1)$-dimensional metric depending 
only on the $(p+1)$-dimensional coordinates $x^{\mu}$, 
$\gamma_{ij}(\Ysp)$ is the $(D-p-1)$-dimensional metric depending 
only on the $(D-p-1)$-dimensional coordinates $y^i$, and 
we assume that the parameters $a$, 
$b$ in the metric (\ref{p:metric:Eq}) are given by  
\Eq{
a=-\frac{4\left(D-p-3\right)}{N\left(D-2\right)},~~~~~~~~~
b=\frac{4\left(p+1\right)}{N\left(D-2\right)}.
 \label{ap:parameter:Eq}
}

The dynamical $p$-brane solutions are characterized by 
function, $h$, depending 
on the $(D-p-1)$-dimensional coordinates $y^i$ transverse to the 
corresponding brane as well as 
the $(p+1)$-dimensional world-volume coordinate $x^{\mu}$\,. 

The expression for the field strength $F_{\left(p+2\right)}$, 
 and scalar field $\phi$ is given by 
\Eqrsubl{ap:ansatz:Eq}{
\e^{\phi}&=&h^{2\epsilon c/N}\,,
  \label{ap:dilaton:Eq}\\
F_{\left(p+2\right)}&=&\frac{2}{\sqrt{N}}\,
d\left[h^{-1}(x, y)\right]\wedge
\sqrt{-q}\,dx^0\,\wedge\,dx^1\wedge\,\cdots\,\wedge\,dx^p\,,
  \label{ap:strength:Eq}
}
where $q$ is the determinant of the metric $q_{\mu\nu}$\,.

Let us first consider the gauge field Eq.~(\ref{ap:gauge:Eq}).
Using the assumptions (\ref{ap:metric:Eq})
and (\ref{ap:ansatz:Eq}), we have 
\Eq{
\pd_\mu \pd_i h=0\,,~~~~~~\lap_{\Ysp}h=0\,,
\label{ap:gauge2:Eq}
 }
where $\lap_{\Ysp}$ is the Laplace operator 
constructed from the metric 
$\gamma_{ij}(\Ysp)$\,.

From Eq.~(\ref{ap:gauge2:Eq})\,, 
the function $h$ can be expressed as 
\Eq{
h(x, y)=h_0(x)+h_1(y)\,,~~~~~~\lap_{\Ysp}h_1=0\,.
  \label{ap:h:Eq}  
}

Using the assumptions (\ref{ap:metric:Eq}) and 
(\ref{ap:ansatz:Eq}), the Einstein equations are reduced to
\Eqrsubl{ap:cEin:Eq}{
&&\hspace{-1.cm}R_{\mu\nu}(\Xsp)-\frac{4}{N}h^{-1}D_{\mu}D_{\nu}h_0
+\frac{2}{N}\left(1-\frac{4}{N}\right)
h^{-2}\pd_{\mu}h_0\pd_{\nu}h_0\nn\\
&&~~~~-\frac{a}{2}h^{-2}q_{\mu\nu}\left[
h\lap_{\Xsp}h_0-\left(1-\frac{4}{N}\right)
q^{\rho\sigma}\pd_{\rho}h_0
\pd_{\sigma}h_0\right]=0, 
 \label{ap:cEin-mu:Eq}\\
&&\hspace{-1.0cm}
R_{ij}(\Ysp)-\frac{b}{2}h^{-2+\frac{4}{N}}\gamma_{ij}\left[h\lap_{\Xsp}h_0
-\left(1-\frac{4}{N}\right)q^{\rho\sigma}\pd_{\rho}h_0\pd_{\sigma}h_0
\right]=0,
  \label{ap:cEin-ij:Eq}
}
where we have used (\ref{ap:h:Eq}). 

Substituting ansatz for fields (\ref{ap:ansatz:Eq}), 
metric (\ref{ap:metric:Eq}), and the equation (\ref{ap:h:Eq}) 
into the equation of motion 
for scalar field (\ref{ap:scalar:Eq}), we find 
\Eq{
\frac{\epsilon c}{N}h^{-b-2+\frac{4}{N}}\left[
h\lap_{\Xsp}h_0-\left(1-\frac{4}{N}\right)q^{\rho\sigma}\pd_{\rho}h_0
\pd_{\sigma}h_0\right]=0.
  \label{ap:scalar-e:Eq}
}
Hence if we assume $c\ne 0$, and 
\Eqrsubl{ap:fields:Eq}{
&&R_{\mu\nu}(\Xsp)=0\,,~~~~~R_{ij}(\Ysp)=0\,,~~~~~N=4\,,\\
&&D_{\mu}D_{\nu}h_0=0\,,~~~~~\lap_{\Ysp}h_1=0\,,
}
the field equations are solved with the condition 
(\ref{ap:h:Eq}).

Let us consider the case in more detail
\Eq{
q_{\mu\nu}(\Xsp)=\eta_{\mu\nu},~~~~~\gamma_{ij}(\Ysp)=\delta_{ij}\,,
 \label{ap:flat:Eq}
 }
where $\eta_{\mu\nu}$ is the $(p+1)$-dimensional 
Minkowski metric, 
 $\delta_{ij}$ is $(D-p-1)$-dimensional Euclidean metrics,
respectively. 
The solution for the functions $h_0$ and $h_1$ can be obtained
explicitly as (\ref{pv:warp:Eq})
\cite{Binetruy:2007tu, Binetruy:2008ev, Maeda:2009zi, 
Uzawa:2010zza, Minamitsuji:2010kb, Minamitsuji:2010uz, Uzawa:2014dra}.

Now we introduce a new time coordinate $\tau$ by
\Eq{
\frac{\tau}{\tau_0}=
\left(c_0t+\bar{c}\right)^{\frac{D+p-1}{2(D-2)}}\,,~~~~~
\tau_0=\frac{2\left(D-2\right)}{c_0\left(D+p-1\right)}\,,
}
where we have assumed $c_0>0$, $c_A=0~(A=1, 2, \cdots , p)$ for simplicity.
Then, the $D$-dimensional metric 
(\ref{ap:metric:Eq}) 
is given by
\Eqr{ 
ds^2&=&\left[1+\left(\frac{\tau}{\tau_0}\right)^{-\frac{2(D-2)}
{D+p-1}}h_1
\right]^{-\frac{D-p-3}{D-2}}
\left[-d\tau^2+\left(\frac{\tau}{\tau_0}\right)^
{-\frac{2(D-p-3)}{D+p-1}}
\delta_{AB}dx^Adx^B
\right.\nn\\
&&\left.\hspace{-1cm}
+\left\{1+\left(\frac{\tau}{\tau_0}\right)^
{-\frac{2(D-2)}{D+p-1}}h_1\right\}
\left(\frac{\tau}{\tau_0}\right)^{\frac{2(p+1)}{D+p-1}}
\gamma_{ij}dy^idy^j\right],
 \label{ap:cos:Eq}
 }
where the metric $\delta_{AB}$ is the spatial part of the $p$-dimensional
Minkowski metric $\eta_{\mu\nu}$.
If we set $h_1=0$, the scale factor of the $p$-dimensional space
is proportional to $\tau^{-\frac{D-p-3}{D+p-1}}$, 
while that for the remaining
$(D-p-1)$-dimensional space is proportional to 
$\tau^{\frac{p+1}{D+p-1}}$.
Thus, in the limit when the terms with $h_1$ are negligible, 
which is realized in the limit $\tau\to\infty$, we have
a Kaluza-Klein-type dynamical solution \cite{Binetruy:2007tu, 
Binetruy:2008ev, Maeda:2009zi, 
Uzawa:2010zza, Minamitsuji:2010kb, Minamitsuji:2010uz, Uzawa:2014dra}. 

Before closing this section, we discuss the dynamical $p$-brane 
solution on the exceptional case of $c=0$\,. 
For $c=0$, the scalar field becomes constant because of the ansatz 
(\ref{ap:ansatz:Eq}), and the equation of motion for the scalar field 
(\ref{ap:scalar:Eq}) is automatically satisfied. The Einstein
equations thus reduce to
\Eqrsubl{ap:fields2:Eq}{
&&R_{\mu\nu}(\Xsp)=0\,,~~~~~R_{ij}(\Ysp)=\frac{a}{2}(p+1)
\,\Lambda\,\gamma_{ij}(\Ysp)\,,~~~~~N=4\,,\\
&&D_{\mu}D_{\nu}h_0=\Lambda\,q_{\mu\nu}(\Xsp)\,,~~~~~\lap_{\Ysp}h_1=0\,,
}
where $\Lambda$ is a constant. The transverse space to the $p$-brane is 
not Ricci flat, but the Einstein space if $\Lambda\ne 0$\,,  If
$q_{\mu\nu}(\Xsp)=\eta_{\mu\nu}$, the function $h_0$ is no longer 
linear but quadratic in the coordinates $x^{\mu}$ 
\cite{Binetruy:2007tu, Binetruy:2008ev, Maeda:2009zi, Uzawa:2010zza}:
\Eq{
h_0(x)=\frac{\Lambda}{2}x^{\mu}x_{\mu}+c_\mu x^\mu + \bar{c}\,,
}
where $c_\mu$ and $\bar{c}$ are constant parameters.

\section{Global structure of static $p$-branes}
\label{sec:static} 
In this Appendix, we give the geometrical structure of a static 
$p$-brane. 
Let us discuss the geodesics for a static $p$-brane solution. 
In general, the metric of $D$-dimensional spacetime for 
a static $p$-brane is given by 
\Eqrsubl{a:bg-s:Eq}{
&&ds^2=h^a(r)\eta_{\mu\nu}(\Xsp)dx^{\mu}dx^{\nu}+h^b(r)\delta_{ij}
(\Ysp)dy^idy^j\,,
 \label{a:metric-s:Eq}\\
&&\gamma_{ij}(\Ysp)dy^idy^j=dr^2+r^2w_{mn}dz^mdz^n\,,\\
&&h_1(r)=1+\frac{M}{r^{D-p-3}}\,, 
  \label{a:h-s:Eq}
}
where $w_{mn}$ is the metric of the $(D-p-2)$-dimensional sphere. 
The behavior of the function $h$ is classified into three classes depending 
on the dimensions of the $p$-brane, which is given by
\Eq{
({\rm i})~p < D-3\,,~~~~({\rm ii})~p=D-3\,,~~~~({\rm iii})~p=D-2\,.
}
Here, the function $h$ for the $(D-2)$-brane 
is given by the linear function of $r$. 
Then, the $D$-dimensional spacetime is not asymptotically flat. 
In string theory, this type 
of solution is classified as a "massive brane". 
We will review it in Appendix~\ref{sec:b}\,.  
For the $(D-3)$-brane, the harmonic function $h$ diverges both 
at infinity and near the $(D-3)$-branes.
Since there is no regular spacetime region near the $(D-3)$-branes, such
solutions are not physically relevant. 
In this Appendix, we will discuss below only in the case of $p<D-3$\,. 

The geodesic equation (\ref{ge:ge-r:Eq}) in the static $p$-brane 
becomes 
\Eq{
\ddot{r}-\frac{\left(a+b\right)\left(D-p-3\right)}{2r}\,\dot{r}^2=0\,.
}
This equation can be solved, to get 
\Eqrsubl{ge:r0:Eq}{
r(s)&=&\alpha_2\left[(1-\alpha)s +\alpha_1\right]^{\frac{1}{1-\alpha}}\,,
~~~~{\rm for}~~~\left(a+b\right)\left(D-p-3\right)\ne 2\,,\\
r(s)&=&\alpha_4\,\e^{\alpha_3\,s}\hspace{3.4cm}{\rm for}~~~
\left(a+b\right)\left(D-p-3\right)=2\,,
}
where $\alpha_i~(i=1,\cdots, 4)$ are constants, and $\alpha$ is defined by
\Eq{
\alpha=\frac{1}{2}\left(a+b\right)\left(D-p-3\right)\,.
    \label{ge:alpha:Eq}
}
The singularity is located at $r=0$. 
For $\alpha\ne 1$\,, the spacetime described
by (\ref{p:metric:Eq}) is geodesically incomplete because a 
null geodesic reaches it in finite $s$. 
Although the null geodesics never reach 
null singularity 
in the case of $\alpha=1$ and $N=4$\,, we have to set $D>11$\,. 
Hence, there is 
no solution for $\alpha=1$ in string or supergravity theory.

For $c_0=0$ and $\alpha\ne 1$ in the limit $r\rightarrow 0$, 
the radial null geodesic equation (\ref{ge:ge-r:Eq}) gives 
\Eq{
t(s)=\pm \,\frac{\left(D-p-3-2\alpha\right)
\alpha_2^{\frac{D-p-5}{2(1-\alpha)}}M^{1/2}}{a(D-p-3)(D-p-5)}\,
\left[\left(1-\alpha\right)s 
+\alpha_1\right]^{-\frac{D-p-5}{2(1-\alpha)}}+t_0\,,
}
for $D-p-3\ne 0$\,, $D-p-5\ne 0$\,, and 
\Eq{
t(s)=t_1\,\ln \left[\left(1-\alpha\right)s +\alpha_1\right]
  +t_0\,,
}
for $D-p-5=0$\,. 
Here, $\alpha_1$ is given in (\ref{ge:t0:Eq}), $t_0$, $t_1$ are 
constants, and $\alpha$ is defined by (\ref{ge:alpha:Eq}). 
As $r\rightarrow 0$ in the $p$-brane ($p\le D-5$)\,, 
$t\rightarrow\pm\infty$\,. On the other hand, for $D-p-5<0$,  
$t$ becomes finite value in the limit $r\rightarrow 0$\,. 
Then the static $p$-brane solution has a null singularity at $r=0$ for 
$D-p-5\ge 0$\,, while there is a timelike singularity at $r=0$ in the 
case of $D-p-5<0$\,.

\begin{figure}[h]
 \begin{center}
\includegraphics[keepaspectratio, scale=0.31, angle=-90]{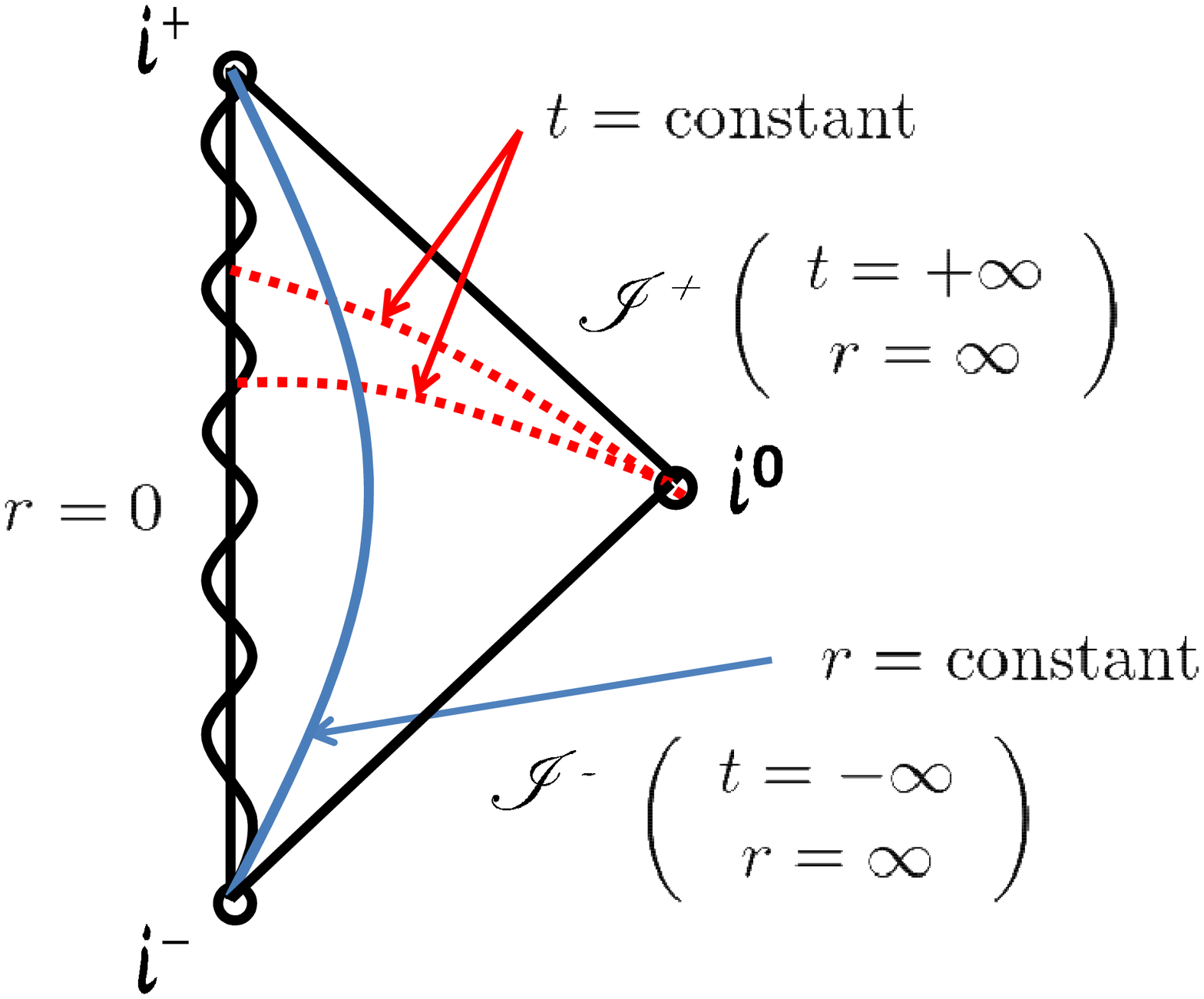}
\hskip 0.5cm
\includegraphics[keepaspectratio, scale=0.31, angle=-90]{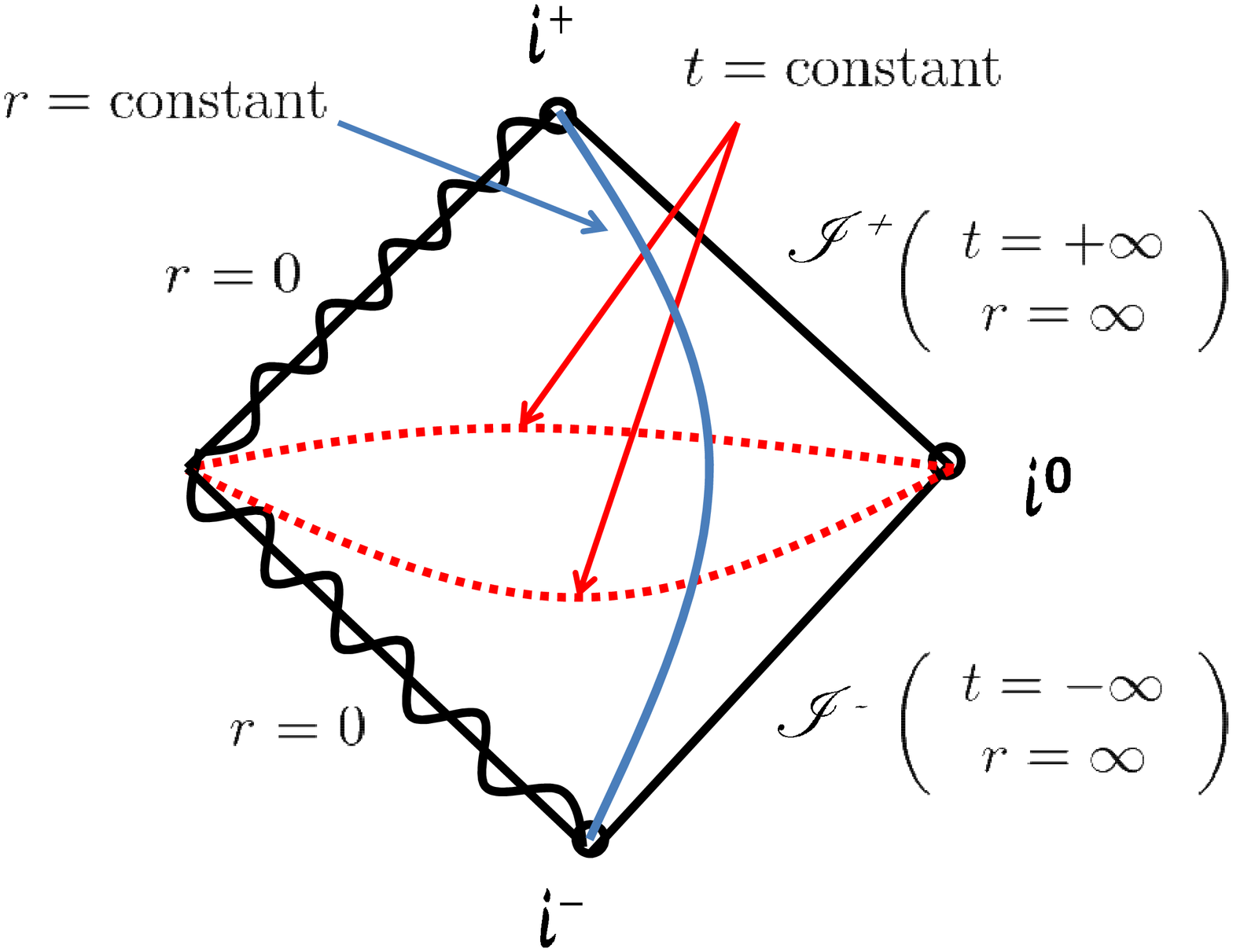}
\\
  \caption{\baselineskip 14pt 
The Penrose diagram of the $D$-dimensional spacetime 
(\ref{g:h:Eq}) with $c_0=0$\,, 
 $N=4$\,, $D-p-3=1$ is depicted in the left panel. 
The wavy line at $r=0$ is the timelike singularity. 
The right panel shows the Penrose diagram for the 
solution (\ref{g:h:Eq}) with $c_0=0$\,, $N=4$\,, $D-p-3\ge 2$\,. 
The wavy lines at $r=0$ are 
the null singularities. }
  \label{fig:sp}
 \end{center}
\end{figure}

The global structure of the $p$-brane spacetime 
(\ref{ge:bg:Eq}) with $c_0=0$ is similar to the 
extreme Reissner-Nordstr\"om solution. 
This global structure is described by the Penrose diagram 
as shown in Fig.~\ref{fig:sp}, in which the
angular coordinates on ${\rm S}^{D-p-2}$ and also the $p$-dimensional 
world-volume coordinates have been suppressed. 


\section{Massive branes}
\label{sec:b} 

In this appendix, we consider the case $D-p-3=-1$ and 
$d_\Ysp=0$ in the 
dynamical $p$-brane solution (\ref{pv:smear:Eq})\,, 
which corresponds to the massive brane in the case of $D=10$, $N=4$ 
\cite{Bergshoeff:1996ui, Bergshoeff:1997ak}.
These solutions generally involve a function that 
is harmonic on the $(D-p-1)$-dimensional space transverse to the 
$(p+1)$-dimensional world-volume of the $p$-brane. 
For $D-p-3>0$, the transverse space has dimension three or greater.  
Thus, there exist harmonic functions that are constant at infinity. 
However, for $D-p-3<0$, the transverse space to $p$-brane has 
dimension two or less. Then, the asymptotic properties are therefore 
qualitatively different. Hence, we pay little attention so far 
to the case of $D-p-3<0$. 

For $D-p-3=-1$ and $N=4$, 
the function of $h$ in the $D$-dimensional metric 
(\ref{p:metric:Eq}) becomes 
\Eq{
h(t, y)=h_0(t)+h_1(y)\,,~~~~h_0(t)=c_0\,t+\bar{c}\,,~~~~
h_1(y)=1+\hat{c}\,y\,, 
    \label{b:h:Eq}
}
where $c_0$\,, $\bar{c}$\,, and $\hat{c}$ are constants\,. 

We consider the Killing symmetry of the $p$-brane background with the case of 
$D-p-3=-1$. 
The solution for the Killing equation can be written by 
\Eq{
\xi=\xi^M\frac{\pd}{\pd x^M}=k^{\mu}\frac{\pd}{\pd x^\mu}
+l\,\frac{\pd}{\pd y}\,,
}
where $k^{\mu}$, $l$ are constant vectors. These vectors obey
\Eq{
k^\mu c_\mu+l\,\hat{c}=0\,.
}
We can also find another Killing vector field $\xi$\,,
\Eq{
\xi=\xi^M\frac{\pd}{\pd x^M}=l\,\frac{\pd}{\pd t}
+k^a\,\frac{\pd}{\pd x^a}-c_0\,\frac{\pd}{\pd y}\,,
}
where the coordinate $x^a~(a=1, 2, \cdots , p)$ 
denotes the spatial part of the 
$(p+1)$-dimensional Minkowski spacetime. 
Since the Killing vector has the norm
\Eq{
g_{MN}\xi^M\xi^N=h^a\left[-l^2+\delta_{ab}\,k^ak^b+hc_0^2\right],
}
the Killing vector is timelike in the case of 
$0<h<\left(l^2-\delta_{ab}\,k^ak^b\right)c_0^{-2}$\,.
If $h>\left(l^2-\delta_{ab}\,k^ak^b\right)c_0^{-2}$\,, 
the Killing vector becomes spacelike. For 
$h=\left(l^2-\delta_{ab}\,k^ak^b\right)c_0^{-2}$\,, 
it is null \cite{Chen:2005jp}. 




\begin{thebibliography}{99}

\bibitem{Horne:1993sy}
  J.~H.~Horne and G.~T.~Horowitz,
  ``Cosmic censorship and the dilaton,''
  Phys.\ Rev.\ D {\bf 48} (1993) R5457
  [hep-th/9307177].

\bibitem{Gibbons:2005rt}
  G.~W.~Gibbons, H.~Lu and C.~N.~Pope,
  ``Brane worlds in collision,''
  Phys.\ Rev.\ Lett.\  {\bf 94} (2005) 131602
  [arXiv:hep-th/0501117].

\bibitem{Chen:2005jp}
  W.~Chen, Z.~-W.~Chong, G.~W.~Gibbons, H.~Lu and C.~N.~Pope,
  ``Horava-Witten stability: Eppur si muove,''
  Nucl.\ Phys.\ B {\bf 732} (2006) 118
  [hep-th/0502077].

\bibitem{Kodama:2005fz}
  H.~Kodama and K.~Uzawa,
  ``Moduli instability in warped compactifications of the type IIB
  supergravity,''
  JHEP {\bf 0507} (2005) 061
  [arXiv:hep-th/0504193].

\bibitem{Kodama:2005cz}
  H.~Kodama and K.~Uzawa,
  ``Comments on the four-dimensional effective theory for warped
  compactification,''
  JHEP {\bf 0603} (2006) 053
  [arXiv:hep-th/0512104].

\bibitem{Kodama:2006ay}
  H.~Kodama and K.~Uzawa,
  ``Moduli instability in warped compactification,''
  hep-th/0601100.

\bibitem{Binetruy:2007tu}
  P.~Binetruy, M.~Sasaki and K.~Uzawa,
  ``Dynamical D4-D8 and D3-D7 branes in supergravity,''
  Phys.\ Rev.\  D {\bf 80} (2009) 026001
  [arXiv:0712.3615 [hep-th]].

\bibitem{Binetruy:2008ev}
  P.~Binetruy, M.~Sasaki and K.~Uzawa,
  ``Dynamical solution of supergravity,''
  arXiv:0801.3507 [hep-th].

\bibitem{Maeda:2009tq}
  K.~i.~Maeda, N.~Ohta, M.~Tanabe and R.~Wakebe,
  ``Supersymmetric Intersecting Branes in Time-dependent Backgrounds,''
  JHEP {\bf 0906} (2009) 036
  [arXiv:0903.3298 [hep-th]].
  
\bibitem{Maeda:2009zi}
  K.~i.~Maeda, N.~Ohta and K.~Uzawa,
  ``Dynamics of intersecting brane systems -- Classification and their
  applications --,''
  JHEP {\bf 0906} (2009) 051
  [arXiv:0903.5483 [hep-th]].

\bibitem{Gibbons:2009dr}
  G.~W.~Gibbons and K.~i.~Maeda,
  ``Black Holes in an Expanding Universe,''
  Phys.\ Rev.\ Lett.\  {\bf 104} (2010) 131101
  [arXiv:0912.2809 [gr-qc]].

\bibitem{Maeda:2009ds}
  K.~i.~Maeda and M.~Nozawa,
  ``Black Hole in the Expanding Universe from Intersecting Branes,''
  Phys.\ Rev.\ D {\bf 81} (2010) 044017
  [arXiv:0912.2811 [hep-th]].
  
\bibitem{Uzawa:2010zza}
  K.~Uzawa,
  ``Dynamical intersecting brane solutions of supergravity,''
  AIP Conf.\ Proc.\  {\bf 1200} (2010) 541.

\bibitem{Maeda:2010ja}
  K.~i.~Maeda and M.~Nozawa,
  ``Black Hole in the Expanding Universe with Arbitrary Power-Law Expansion,''
  Phys.\ Rev.\ D {\bf 81} (2010) 124038
  [arXiv:1003.2849 [gr-qc]].
  
\bibitem{Minamitsuji:2010fp}
  M.~Minamitsuji, N.~Ohta and K.~Uzawa,
  ``Dynamical solutions in the 3-Form Field Background in the
  Nishino-Salam-Sezgin Model,''
  Phys.\ Rev.\  D {\bf 81} (2010) 126005
  [arXiv:1003.5967 [hep-th]].

\bibitem{Maeda:2010aj}
  K.~i.~Maeda, M.~Minamitsuji, N.~Ohta and K.~Uzawa,
  ``Dynamical $p$-branes with a cosmological constant,''
  Phys.\ Rev.\  D {\bf 82} (2010) 046007
  [arXiv:1006.2306 [hep-th]].

\bibitem{Minamitsuji:2010kb}
  M.~Minamitsuji, N.~Ohta and K.~Uzawa,
  ``Cosmological intersecting brane solutions,''
  Phys.\ Rev.\  D {\bf 82} (2010) 086002
  [arXiv:1007.1762 [hep-th]].

\bibitem{Uzawa:2010zz}
  K.~Uzawa,
  ``Cosmological intersecting brane solutions in string theory,''
  J.\ Phys.\ Conf.\ Ser.\  {\bf 259} (2010) 012032.

\bibitem{Nozawa:2010zg}
  M.~Nozawa and K.~i.~Maeda,
  ``Cosmological rotating black holes in five-dimensional fake supergravity,''
  Phys.\ Rev.\ D {\bf 83} (2011) 024018
  [arXiv:1009.3688 [hep-th]].
  
\bibitem{Minamitsuji:2010uz}
  M.~Minamitsuji and K.~Uzawa,
  ``Cosmology in $p$-brane systems,''
  Phys.\ Rev.\  D {\bf 83} (2011) 086002
  [arXiv:1011.2376 [hep-th]].

\bibitem{Maeda:2011sh}
  K.~i.~Maeda and M.~Nozawa,
  ``Black hole solutions in string theory,''
  Prog.\ Theor.\ Phys.\ Suppl.\  {\bf 189} (2011) 310
  [arXiv:1104.1849 [hep-th]].
  
\bibitem{Minamitsuji:2011jt}
  M.~Minamitsuji and K.~Uzawa,
  ``Dynamics of partially localized brane systems,''
  Phys.\ Rev.\ D {\bf 84} (2011) 126006
  [arXiv:1109.1415 [hep-th]].

\bibitem{Maeda:2012xb}
  K.~-i.~Maeda and K.~Uzawa,
  ``Dynamical brane with angles: Collision of the universes,''
  Phys.\ Rev.\ D {\bf 85} (2012) 086004
  [arXiv:1201.3213 [hep-th]].

\bibitem{Blaback:2012mu}
  J.~Bl\aa b\"ack, B.~Janssen, T.~Van Riet and B.~Vercnocke,
  ``Fractional branes, warped compactifications and backreacted orientifold planes,''
  JHEP {\bf 1210} (2012) 139
  [arXiv:1207.0814 [hep-th]].

\bibitem{Minamitsuji:2012if}
  M.~Minamitsuji and K.~Uzawa,
  ``Cosmological brane systems in warped spacetime,''
  Phys.\  Rev.\ D {\bf 87} (2013) 046010
  [arXiv:1207.4334 [hep-th]].

\bibitem{Uzawa:2013koa}
  K.~Uzawa and K.~Yoshida,
  ``Dynamical Lifshitz-type solutions and aging phenomena,''
  Phys.\ Rev.\ D {\bf 87} (2013) 106003
  [arXiv:1302.5224 [hep-th]].

\bibitem{Uzawa:2013msa}
  K.~Uzawa and K.~Yoshida,
  ``Dynamical F-strings intersecting D2-branes in type IIA supergravity,''
  Phys.\ Rev.\ D {\bf 88} (2013) 066005
  [arXiv:1307.3093].

\bibitem{Blaback:2013taa}
  J.~Bl\aa b\"ack, B.~Janssen, T.~Van Riet and B.~Vercnocke,
  ``BPS domain walls from backreacted orientifolds,''
  JHEP {\bf 1405} (2014) 040
  [arXiv:1312.6125 [hep-th]].
  
\bibitem{Uzawa:2014kka}
  K.~Uzawa and K.~Yoshida,
  ``Probe brane dynamics on cosmological brane backgrounds,''
  Phys.\ Lett.\ B {\bf 738} (2014) 493
  [arXiv:1401.3664 [hep-th]].

\bibitem{Uzawa:2014dra} 
  K.~Uzawa,
  ``Colliding $p$-branes in the dynamical intersecting brane system,''
  Phys.\ Rev.\ D {\bf 90}, 025024 (2014)
  [arXiv:1407.7406 [hep-th]].

\bibitem{Kastor:1992nn}
  D.~Kastor and J.~H.~Traschen,
  ``Cosmological multi - black hole solutions,''
  Phys.\ Rev.\  D {\bf 47} (1993) 5370
  [arXiv:hep-th/9212035].

\bibitem{Maki:1992tq}
  T.~Maki and K.~Shiraishi,
  ``Multi - black hole solutions in cosmological Einstein-Maxwell dilaton theory,''
  Class.\ Quant.\ Grav.\  {\bf 10} (1993) 2171
  [arXiv:1403.1320 [gr-qc]].

\bibitem{Brill:1993tm}
  D.~R.~Brill, G.~T.~Horowitz, D.~Kastor and J.~H.~Traschen,
  ``Testing cosmic censorship with black hole collisions,''
  Phys.\ Rev.\ D {\bf 49} (1994) 840
  [gr-qc/9307014].

\bibitem{Penrose:1964wq}
  R.~Penrose,
  ``Gravitational collapse and space-time singularities,''
  Phys.\ Rev.\ Lett.\  {\bf 14} (1965) 57.
  
\bibitem{Penrose:1969pc}
  R.~Penrose,
  ``Gravitational collapse: The role of general relativity,''
  Riv.\ Nuovo Cim.\  {\bf 1} (1969) 252
   [Gen.\ Rel.\ Grav.\  {\bf 34} (2002) 1141].

\bibitem{Maki:1994wc}
  T.~Maki and K.~Shiraishi,
  ``Exact solutions for gravitational collapse with a dilaton field in arbitrary dimensions,''
  Class.\ Quant.\ Grav.\  {\bf 12} (1995) 159
  [arXiv:1504.03062 [gr-qc]].

\bibitem{Johnson:1999qt}
  C.~V.~Johnson, A.~W.~Peet and J.~Polchinski,
  ``Gauge theory and the excision of repulson singularities,''
  Phys.\ Rev.\  D {\bf 61} (2000) 086001
  [arXiv:hep-th/9911161].

\bibitem{Jarv:2000zv}
  L.~Jarv and C.~V.~Johnson,
  ``Orientifolds, M theory, and the ABCD's of the enhancon,''
  Phys.\ Rev.\ D {\bf 62} (2000) 126010
  [hep-th/0002244].
  
\bibitem{Peet:1999nh}
  A.~W.~Peet,
  ``Excision of `repulson' singularities: A Space-time result 
  and its gauge theory analog,''
  hep-th/0003251.

\bibitem{Johnson:2000ch}
  C.~V.~Johnson,
  ``D-brane primer,''
  hep-th/0007170.

\bibitem{Yamaguchi:2001yd}
  S.~Yamaguchi,
  ``Enhancon and resolution of singularity,''
  gr-qc/0108084.

\bibitem{Bergshoeff:1996ui}
  E.~Bergshoeff, M.~de Roo, M.~B.~Green, G.~Papadopoulos and P.~K.~Townsend,
  ``Duality of type II 7 branes and 8 branes,''
  Nucl.\ Phys.\ B {\bf 470} (1996) 113
  [hep-th/9601150].

\bibitem{Bergshoeff:1997ak}
  E.~Bergshoeff, Y.~Lozano and T.~Ortin,
  ``Massive branes,''
  Nucl.\ Phys.\ B {\bf 518} (1998) 363
  [hep-th/9712115].

\end{thebibliography}
\end{document}